\documentclass[%
 preprint, 
superscriptaddress,
 twocolumn,
 amsmath,amssymb,
 aps, physrev,
]{revtex4-2}

\usepackage{graphicx}
\usepackage{dcolumn}
\usepackage{bm}
\usepackage{hyperref}
\usepackage{orcidlink}  

\begin{document}

\preprint{DOI: \href{https://doi.org/10.1103/7572-l2zp}{10.1103/7572-l2zp}}

\title{\textbf{Spatial dependence of the break in the energy spectrum of cosmic rays in the new anisotropic diffusion approach} 
}%

\author{V.D. Borisov\,\orcidlink{0009-0005-2862-7133}}
\email[]{borisov.vd.19@physics.msu.ru}
\affiliation{Faculty of Physics, M.V. Lomonosov Moscow State University (MSU),
119991 Moscow, Russia}
\affiliation{Skobeltsyn Institute of Nuclear Physics, Lomonosov Moscow State University (MSU),
119991 Moscow, Russia}

\author{V.O. Yurovsky\,\orcidlink{0009-0008-1031-4499}}
\affiliation{Faculty of Physics, M.V. Lomonosov Moscow State University (MSU),
119991 Moscow, Russia}
\affiliation{Skobeltsyn Institute of Nuclear Physics, Lomonosov Moscow State University (MSU),
119991 Moscow, Russia}

\author{A.I. Peryatinskaya\,\orcidlink{0009-0009-8802-9302}}
\affiliation{Faculty of Physics, M.V. Lomonosov Moscow State University (MSU),
119991 Moscow, Russia}
\affiliation{Skobeltsyn Institute of Nuclear Physics, Lomonosov Moscow State University (MSU),
119991 Moscow, Russia}

\author{I.A. Kudryashov\,\orcidlink{0009-0009-1889-6232}}
\email[]{ilya.kudryashov.85@gmail.com}
\affiliation{Skobeltsyn Institute of Nuclear Physics, Lomonosov Moscow State University (MSU),
119991 Moscow, Russia}




\begin{abstract}
At present, there is no consensus on whether the spectral break in the cosmic-ray flux of all elements around 4 PeV is a general characteristic of the Milky Way or is determined by a combination of factors that significantly affect the energy position of the knee. We argue that considering the anisotropic propagation of cosmic rays within a realistically modeled Galactic magnetic field enables an accurate description of the spectral break without invoking a source-related cutoff energy, and naturally predicts a spatial dependence of this phenomenon. To demonstrate this, we constructed a 3D diffusion propagation model with a diffusion tensor within a two-component magnetic field and determined both the local cosmic-ray spectrum and the spectra for specific regions of the Milky Way. We found that the spectral break is well reproduced by the energy dependence of the diffusion tensor components and appears at different energies in various regions of the Galaxy due to the inclusion of the turbulent component of the magnetic field. Furthermore, the spectral slope is determined by the degree of anisotropy in cosmic-ray diffusion. The model is based on the calculation of the diffusion tensor components using the trajectory method and on the direct solution of the stationary diffusion equation with a fully anisotropic diffusion tensor that includes all nine components in the global Galactic coordinate system, accounting for off-diagonal terms arising from the projection of locally field-aligned diffusion. We calculated the integral flux of diffuse gamma rays in the inner and outer regions of the Galaxy based on the cosmic-ray proton and nuclei spectra obtained within our model, which features a spatially dependent position of the knee. The resulting spectral shape of the diffuse gamma-ray flux is consistent with experimental data from LHAASO and Fermi-LAT.
\end{abstract}

\maketitle


\section{\label{sec:level1}Introduction}

The spectrum of all cosmic-ray (CR) elements in the vicinity of Earth has been measured with high statistical accuracy. The spectral index changes from $\gamma \approx -2.7$ to $\gamma \approx -3.1$ around 4 PeV ~\cite{H_randel_2003}. Despite some discrepancies among different experiments regarding the precise energy position of the spectral break, the fact of the spectral index change is observed across all experiments. The break in the spectrum is generally explained either by the acceleration limit at the CR sources ~\cite{Kachelrie__2019}, typically at the shock waves of supernova remnants, or by specific propagation effects in the interstellar medium. Theoretical estimates of the source spectrum yield a spectral index of $\alpha \sim [-2.0; -2.4]$ ~\cite{H_randel_2003,Zirakashvili2021}, with a cutoff energy in the range of (10–-100 TeV). The steepening of the spectrum in this approach is attributed to diffusive propagation, which is divided into several main approaches: the isotropic diffusion approach, investigated in numerous models ~\cite{Strong1998, Lagutin2001} and long considered the primary one due to its simplicity, and the anisotropic diffusion approach ~\cite{Giacinti_2015, Evoli_2017}, which is computationally more complex but allows for the description of several more complicated effects.

In widely used propagation codes such as ~\cite{Effenberger_2012, Evoli_2017}, the energy dependence of the diffusion coefficients is not computed from first principles but instead is parametrized phenomenologically. Typically, an isotropic or mildly anisotropic diffusion tensor is assumed, with the scalar coefficient expressed as a power law in energy:
\begin{equation}
\label{eq:D}
D(E) = D_0\left(\frac{E}{E_0}\right)^\delta,
\end{equation}
where $D_0$, $E_0$, and $\delta$ are free parameters. These are empirically calibrated to reproduce observational constraints, such as the ratios of secondary to primary cosmic rays (e.g., B/C), and the measured energy spectra of protons and electrons. This leads to the diffusion tensor not being derived from the magnetic field structure or turbulence properties, but instead being adjusted to fit observations without explicitly modeling the microphysics of particle transport.

Moreover, existing models of anisotropic diffusion often suffer from significant limitations. Some do not address the full 3D nature of cosmic-ray propagation, while others, even in three dimensions, adopt coordinate transformations aligned with the large-scale Galactic magnetic field that result in vanishing off-diagonal components of the diffusion tensor. This simplification, while computationally convenient, effectively removes the anisotropic nature of the transport and introduces unphysical correlations between orthogonal spatial directions.

To overcome these limitations, we developed a fully 3D, implicit finite-difference scheme that enables a physically accurate coordinate transformation to the local Galactic magnetic field, preserving the anisotropy of the diffusion tensor throughout the simulation volume. Our method yields computational run times comparable to those of traditional diffusion models yet offers a significantly improved physical foundation.

We compute the energy dependence of the diffusion tensor components via first-principles simulations of test-particle trajectories in prescribed magnetic field configurations. These include both regular and turbulent components, characterized by a specified correlation length and turbulence power spectrum. This approach ensures that the diffusion coefficients are based on the physical processes of particle scattering and transport and can be directly scaled according to the local regular and turbulent magnetic field properties at each grid node, as described in Secs.~\ref{sec:Propagation} and \ref{sec:Method}."

We present the model of anisotropic CR propagation with a fully anisotropic diffusion tensor that includes all nine components in the global Galactic coordinate system, accounting for off-diagonal terms arising from the projection of locally field-aligned diffusion. The calculation of the diffusion tensor components is performed using the trajectory method. We apply this model to analyze recent experimental data from LHAASO ~\cite{Cao_2023} and Fermi-LAT ~\cite{Zhang_2023}, as well as CR nuclear groups.

Understanding the propagation mechanism of CRs within our Galaxy is critically important for a wide range of astrophysical problems: the accurate study of the evolution of various source populations, the estimation of fluxes of both the leptonic and hadronic CR components, and consequently, the derivation of constraints on the mass of potential dark matter carriers and the description of CR arrival anisotropy, as determined by the IceCube experiment ~\cite{abbasi2024observationcosmicrayanisotropysouthern}. Accurate model parameters can be determined by combining CR nuclear group spectra with experimental data from regions of the Galaxy different from the location of our Solar System. Such experiments cannot be conducted directly; however, the latest ground-based experiments, such as LHAASO and Tibet AS$\gamma$ ~\cite{Amenomori_2021}, along with space-based experiments like Fermi-LAT, allow for the estimation of gamma-ray fluxes in the energy range up to 1 PeV. The primary source of such gamma rays (above 10 TeV) is the decay of $\pi^+, \pi^-$ and $\pi^0$ mesons produced in nuclear interactions between CR nuclei and gas or molecular clouds ~\cite{Ackermann_2012}. Moreover, the number of $\pi^+$ and $\pi^-$ mesons is suppressed relative to $\pi^0$ mesons by approximately a factor of 10. Recent studies on the joint analysis of gamma-ray data from the LHAASO and Fermi-LAT experiments ~\cite{prevotat2024energydependencekneecosmic} indicate the inability to describe the observed experimental data within the framework of a simple diffusion approach and suggest a potential spatial dependence of the CR knee.

We argue that transitioning to a fully anisotropic approach naturally explains the CR knee and demonstrates its spatial dependence as a result of a change in the propagation mechanism. To illustrate this, we have developed a model of anisotropic CR transport in a two-component magnetic field, which enables the estimation of CR concentrations within a given volume of the Galaxy. Additionally, we have implemented a method for calculating high-energy gamma-ray fluxes.

The paper is organized as follows: In Sec. \ref{sec:Propagation}, we describe the fundamental ideas of the approach used to model microscopic and macroscopic transport, as well as the free parameters of the model, such as the correlation length, turbulence spectrum, source distribution, large-scale Galactic magnetic fields, and characteristic parameters of the modeled volume. Section \ref{sec:Method} details the numerical solution method. In Sec. \ref{sec:anis_features}, we model the effects of anisotropic propagation on cosmic-ray spectra. In Sec. \ref{sec:all_elements}, we present the description of element-specific spectra as well as the spectrum of all CR elements. Section \ref{sec:results} compares the computed model gamma-ray fluxes with data from LHAASO and Fermi-LAT in both the inner and outer Galaxy, demonstrating the feasibility of their description within the anisotropic diffusion approach. Finally, we conclude with a summary of the main results in Sec. \ref{sec:conclusion}.

\section{\label{sec:Propagation} Features of CR Propagation}

Galactic CRs are relativistic particles that propagate from their sources to us through interstellar magnetic fields. Several models of the Galactic magnetic field exist, among which JF12 \cite{Jansson_2012} and Unger24\cite{Unger_2024} stand out as relatively simple yet qualitatively effective representations of the large-scale structure of our Galaxy's magnetic field. Many models also assume the presence of a turbulent component. The transport of Galactic cosmic rays (GCRs) in turbulent magnetic fields has been extensively studied ~\cite{Mertsch_2020, Effenberger_2012, kuhlen2023}, and, in particular, a numerical model of GCR transport in a turbulent magnetic field was developed by the authors in ~\cite{Kudryashov2023}.

The transport of ultrarelativistic charged particles in a turbulent magnetic field is highly complex. Over sufficiently long distances, it can be considered diffusive; however, when a "mean" regular field is present, diffusion exhibits a highly anisotropic nature (see, for example, ~\cite{kuhlen2023, Kudryashov2023}). Although at small scales, GCR transport follows either ballistic or superdiffusive behavior, in this study, we adopt the diffusive approximation:

\begin{equation}
\label{eq:D}
\lim_{t \to +\infty} \frac{\left< x_i(t)x_j(t)\right>}{t} = D_{ij},
\end{equation}

where $D_{ij}$ is the diffusion tensor aligned with the local magnetic field.

In this work, we do not consider fragmentation, reacceleration of GCRs, energy losses due to radiation, or interactions with matter.We focus on solving the steady-state problem, which, while not capturing possible time-dependent effects at the highest energies, provides a reliable description of the main spectral features and spatial distribution of CRs within the considered energy range. This leads to the following equation for the CR particle density:

\begin{equation}
\nabla D_{ij}(\mathbf{B(\mathbf{r})},E) \nabla n (E,\mathbf{r}) = q(E,\mathbf{r}),
\label{eq:difusion_eq}
\end{equation}

where $n(E,\mathbf{r})$ is the CR density and $q(E,\mathbf{r})$ is the source density. In this study, we assume that the sources are distributed according to ~\cite{yusifov2004galacticdistributionluminosityfunction} and the magnetic field is modeled using Unger24.

The energy dependence of the diffusion tensor components $D_{\parallel}$ and $D_{\perp}$ is obtained via direct numerical integration of charged particle trajectories in a model turbulent magnetic field. The field includes both regular and turbulent components, with turbulence following a Kolmogorov power spectrum:
\begin{equation}\label{eq:kolmog}
P(k) \propto k^{-5/3}.
\end{equation}

Proton test particles are initialized with isotropic pitch-angle distributions and tracked by numerically solving the relativistic equations of motion under the Lorentz force. The integration is carried out until the mean square displacements along the radial and vertical directions exhibit linear growth in time, indicative of diffusive behavior. The diffusion coefficients are then computed using Eq. (\ref{eq:D}). Although note that correlation length $\lambda_c$ of the magnetic turbulence and the turbulence power spectrum are not explicitly present in the diffusion tensor equations, they are implicitly encoded in the structure of the turbulent field. Specifically, $\lambda_c$ is defined by the spectral distribution of wave modes in the field, with the outer scale of turbulence determining the scale at which particles experience the strongest scattering. Repeating this procedure for different particle energies allows the construction of the energy dependence of the diffusion tensor components.\par
Once the energy dependence of $D_{\parallel}(E)$ and $D_{\perp}(E)$  is computed for a reference configuration (i.e., for a specific value of $B_0$, $\lambda_c$ and $P(k)$), it can be scaled to arbitrary field conditions throughout the Galaxy. In the numerical solution of the steady-state transport equation \ref{eq:difusion_eq}, this scaling is applied at each grid point of the spatial mesh. The local values of the regular and turbulent magnetic field strengths -- defined according to the chosen Galactic magnetic field model (e.g., Unger24) and the assumed spatial distribution of turbulence -- are used to scale the diffusion coefficients accordingly, following the prescription detailed in Sec. \ref{sec:Method}. Further details of the calculation are provided in the Appendix \ref{app:subsecA}.

Figure \ref{fig:D1} illustrates the longitudinal and transverse diffusion coefficients $D_{ij}$ in Eq. (\ref{eq:difusion_eq}) for different magnitudes of the "regular" magnetic field. The solid orange and dashed red lines show the dependence of the parallel diffusion coefficient, while the solid blue and dashed green lines represent the perpendicular diffusion coefficient for two different spatial locations within the modeled volume of the Galaxy, their Cartesian coordinates (in kpc) relative to the Galactic center are as follows: B (-8.2,0.0,0.2) and C (5.4,0.0,-2.0). The change in the power-law behavior occurs at different energies depending on the magnetic field strength.

\begin{figure}[h]
\includegraphics[width=0.5\textwidth]{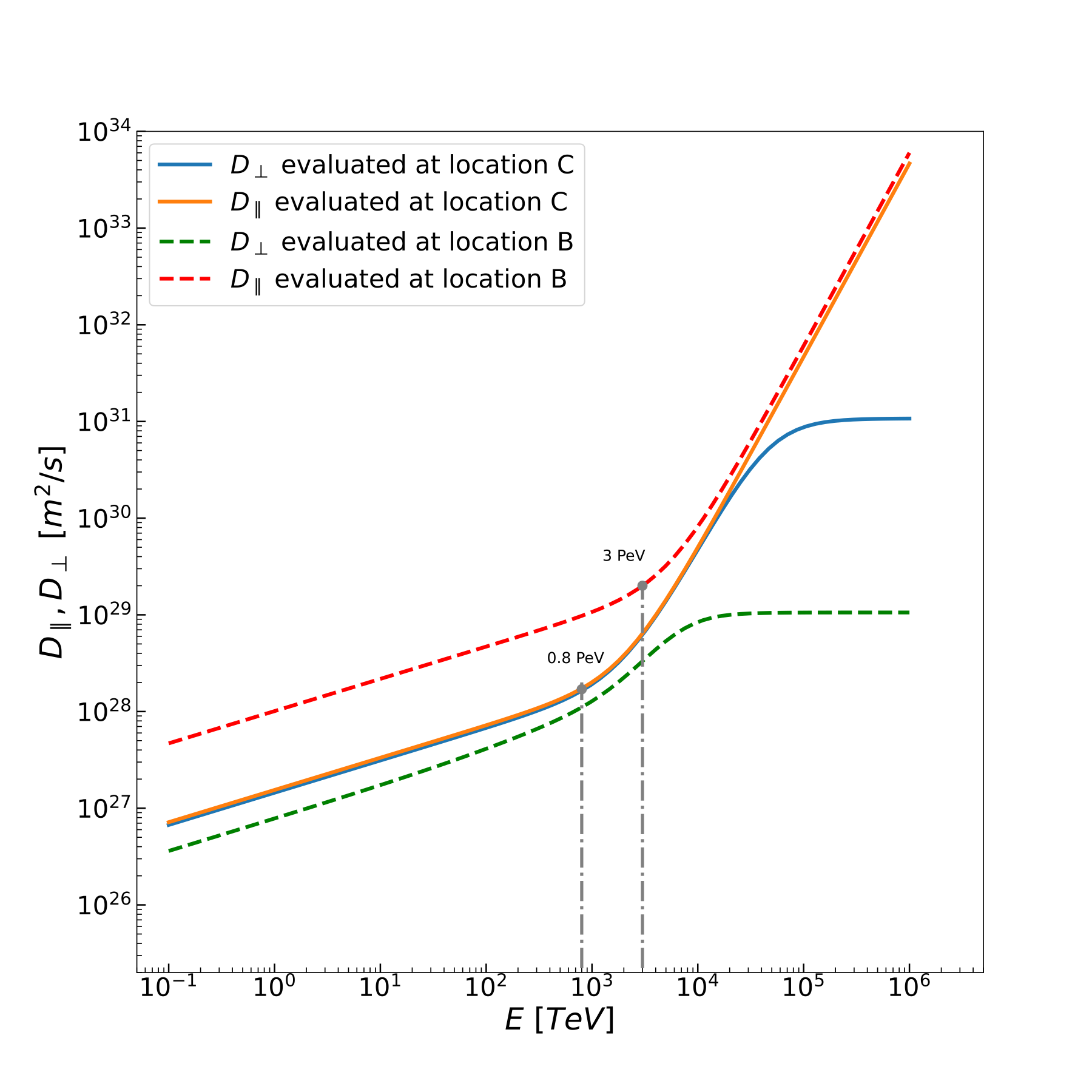}
\caption{\label{fig:D1} Dependence of the longitudinal $D_{\parallel}$ and transverse $D_{\perp}$ components of the diffusion tensor on energy for different magnitudes of the magnetic field modulus.}
\end{figure}

A key region is the energy range from approximately $10^2$ to $10^4$ TeV, where the transport regime transitions. This transition occurs at different energies depending on the magnitude of the regular magnetic field, the turbulence correlation length $\lambda_c$, the turbulence spectrum, and the specific turbulence model considered. We refer to the energy of this transition as the critical point. To the left of the critical point, $D_{\parallel}$ and $D_{\perp}$ increase as $const\cdot E^{1/3}$, where the constant is a normalization factor determined by the local field. This dependence arises due to the strong magnetization of the particles, whose gyroradii $r_g$ are smaller than the characteristic scales of magnetic field inhomogeneities. A unified theoretical description for this regime remains elusive, with various numerical solutions differing in power-law indices but yielding the same critical points. To the right of the critical point, $D_{\parallel}$ and $D_{\perp}$ behave differently: the longitudinal diffusion coefficient $D_{\parallel}$ continues to increase, albeit with a scaling of $const\cdot E^{2}$. Here, the gyroradius becomes large enough that scattering on magnetic inhomogeneities weakens. Meanwhile, $D_{\perp}$ saturates at a constant value, leading to pronounced inhomogeneities in the propagation of charged particles.

\section{Numerical Solution Method} \label{sec:Method}

An important feature of our problem formulation is that the diffusion tensor depends not only on the particle energy, but also on the Galactocentric coordinates due to the spatial dependence of the magnetic field components. The components of the diffusion tensor are calculated by numerically solving the equation of motion of a particle in a magnetic field:

\begin{equation}\label{eq:5}
\frac{d\mathbf{r}}{dt} = \mathbf{v}
\end{equation}

\begin{equation}\label{eq:6}
\frac{d\mathbf{v}}{dt} = \frac{q}{E} \mathbf{v} \times \mathbf{B}
\end{equation}

where $\mathbf{r}$ represents the particle coordinates, $\mathbf{v}$ is the particle velocity, $q$ is the particle charge, $E$ is the particle energy, and $\mathbf{B}$ is the magnetic field.

The solution is performed using the Cash-Karp method, a modification of the fourth-order Runge-Kutta method.

The time step is empirically chosen to accurately model motion in a uniform magnetic field. To verify the correctness of the chosen step, results were compared with those obtained using a tenfold increase in step size, which did not lead to statistically significant deviations.

For the numerical solution of the diffusion equation \ref{eq:difusion_eq}, we use a fully implicit finite-difference scheme, ensuring stability of the solution with respect to the input data. Equation \ref{eq:difusion_eq} is factorized for each given energy and magnetic field strength as follows:

\begin{equation}
\frac{\partial}{\partial x_i} \left( D_{ij}(\mathbf{r}) \frac{\partial f(\mathbf{r})}{\partial x_j} \right) = S(x_j),
\end{equation}

where $S(x_j)$ represents the source concentration at the point $\mathbf{r} = (x_1, x_2, x_3)$.

Further expansion of Derivatives:

\begin{multline}
\frac{\partial}{\partial x_i} \left( D_{ij}(\mathbf{r}) \frac{\partial f(\mathbf{r})}{\partial x_j} \right) =
\frac{\partial D_{ij}(\mathbf{r})}{\partial x_i} \frac{\partial f(\mathbf{r})}{\partial x_j} + \\
D_{ij}(\mathbf{r}) \frac{\partial^2 f(\mathbf{r})}{\partial x_i \partial x_j}
\end{multline}

Recursive substitution:

\begin{equation}
\frac{\partial g(\mathbf{r})}{\partial x_i} \approx \frac{g(\mathbf{r} + x_i h_i) - g(\mathbf{r} - x_i h_i)}{2 h_i},
\end{equation}

where the function $g(\mathbf{r})$ is replaced by the specific tensor coefficient $D_{ij}(\mathbf{r})$.

\begin{figure*}
\includegraphics[width=0.95\textwidth]{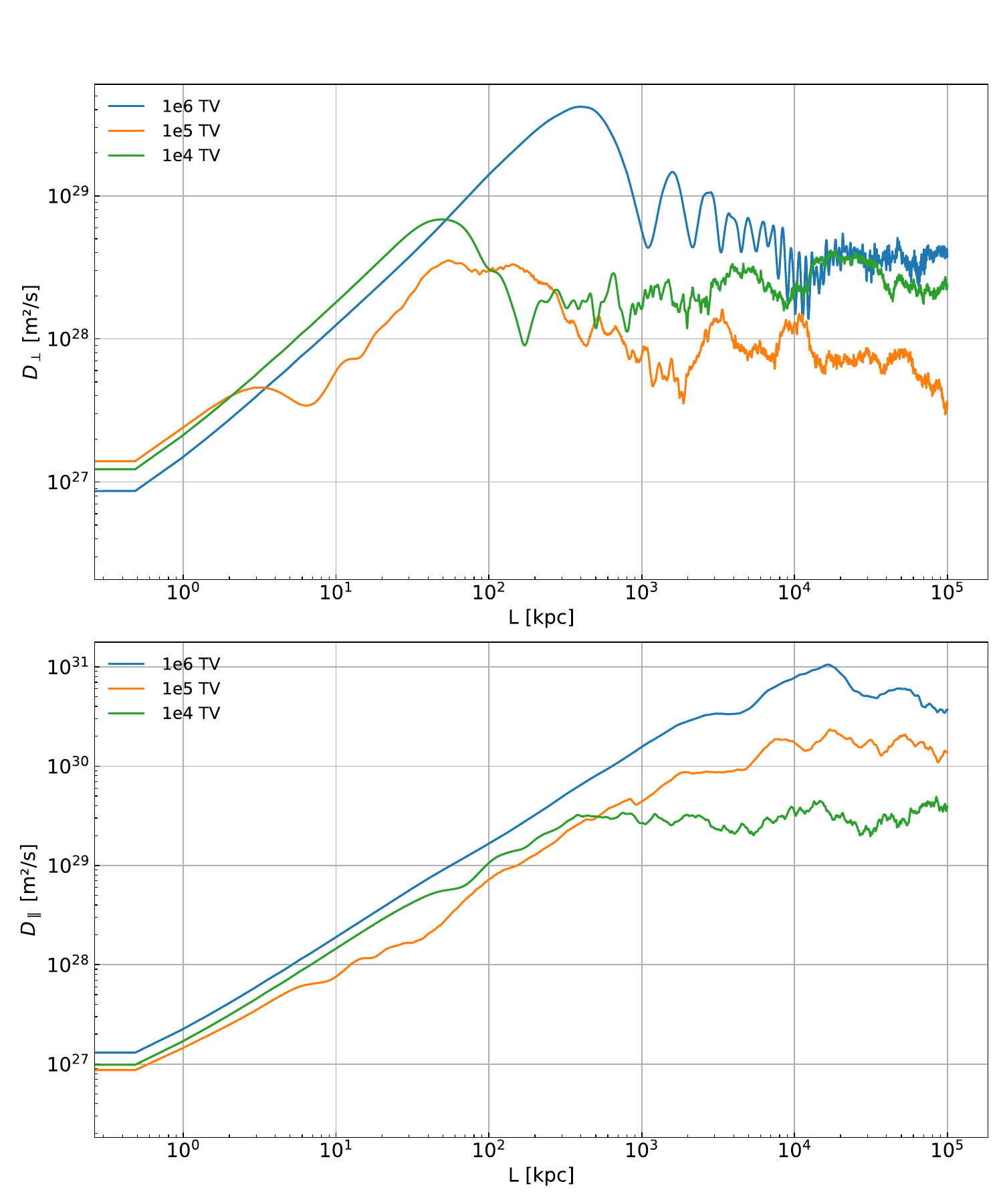}
\caption{\label{fig:path}The characteristic displacement L required for the transition to the diffusive propagation regime. The increase of the diffusion coefficients saturates once the diffusion regime is reached.}
\end{figure*}

The finite-difference grid steps are individually selected for each energy to ensure that the displacement of a charged particle along its trajectory is sufficient for reaching the diffusive transport regime. Figure \ref{fig:path} shows the characteristic displacements associated with the transition to diffusion for both the parallel ($D_\parallel$ and perpendicular ($D_\perp$) diffusion coefficients in the limiting regime of applicability of the diffusion approximation (magnetic rigidity $R = 10^4$--$10^6$ TV).
It should be noted that the transition to diffusion occurs at different displacements for $D_\parallel$ and $D_\perp$ even at the same rigidity. Specifically, $D_\perp$ reaches the diffusive regime earlier than $D_\parallel$, indicating the existence of a mixed propagation regime in this range.

For protons with an energy of approximately 10 PeV, the characteristic displacements (equivalent to the spatial step in the finite-difference scheme) are about 0.2 kpc, increasing to 4–-10 kpc at energies of 100–-1000 PeV. The mixed propagation regime appears over displacements from 0.1 to 0.8 kpc. For comparison, the modeled volume size (e.g., the major axis of the Galactic disk) is about 40 kpc.

Therefore, the diffusion approximation remains fully valid for rigidities up to approximately $R \sim 10^4$ TV, and gradually becomes inadequate in the range $R \sim 10^5$ -- $10^6$ TV where the use of test-particle methods becomes justified. For rigidities below $10^4$ TV, the characteristic displacements are smaller than 0.1 kpc, allowing for a finer grid resolution.

\section{\label{sec:anis_features}Modeling the Influence of Anisotropic Propagation on Cosmic-Ray Spectra}

To evaluate the impact of the energy dependence of the diffusion tensor components -- both parallel (along the magnetic field)
\begin{figure*}[htbp]
\centering
\begin{minipage}{0.45\textwidth}
\centering
\includegraphics[width=\textwidth]{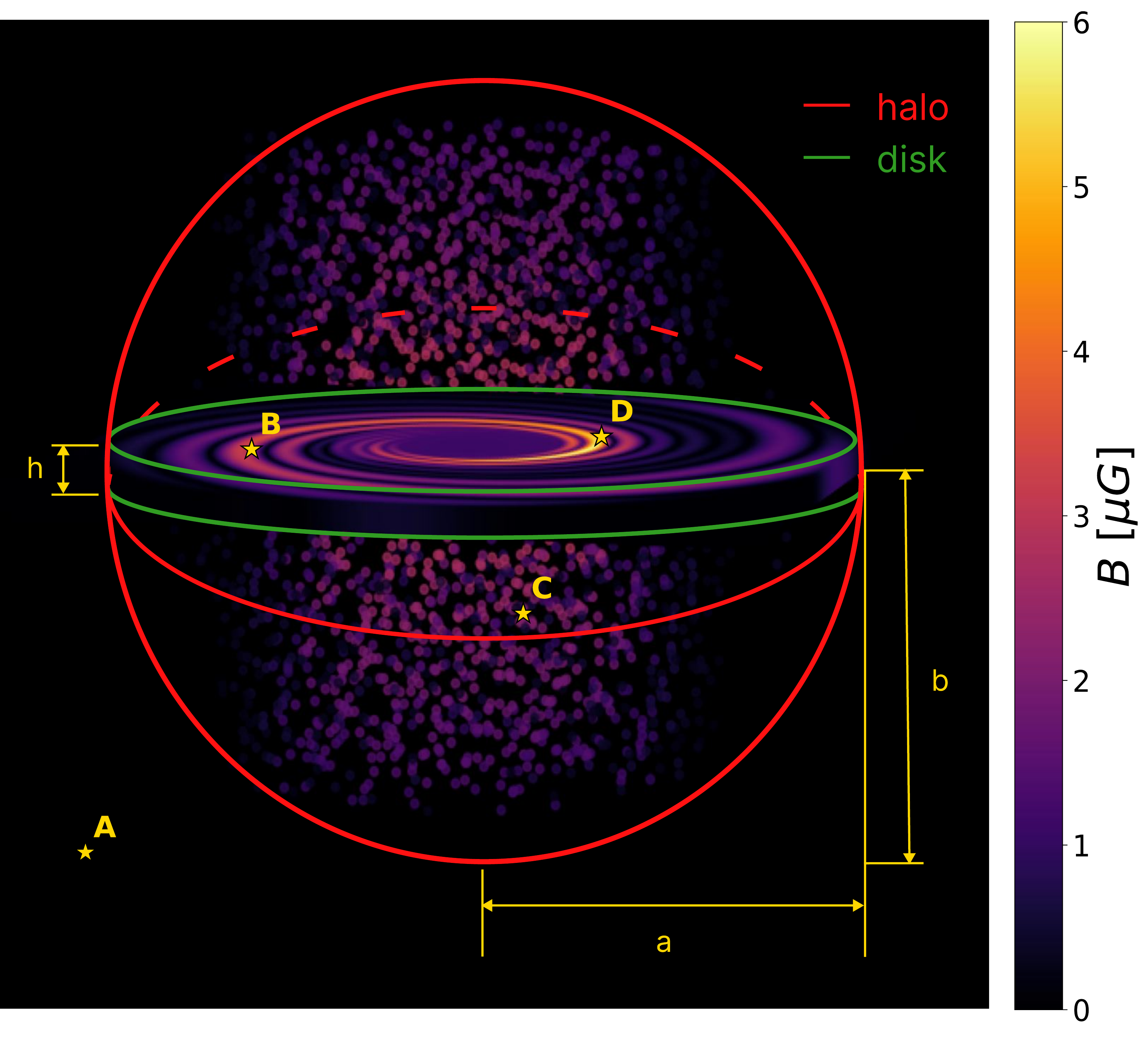}
\caption{Schematic representation of the modeled Galactic volume. a=17 kpc; b=6 kpc; h=0.4 kpc.}
\label{fig:galaxy}
\end{minipage}%
\hfill
\begin{minipage}{0.45\textwidth}
\centering


\includegraphics[width=\textwidth]{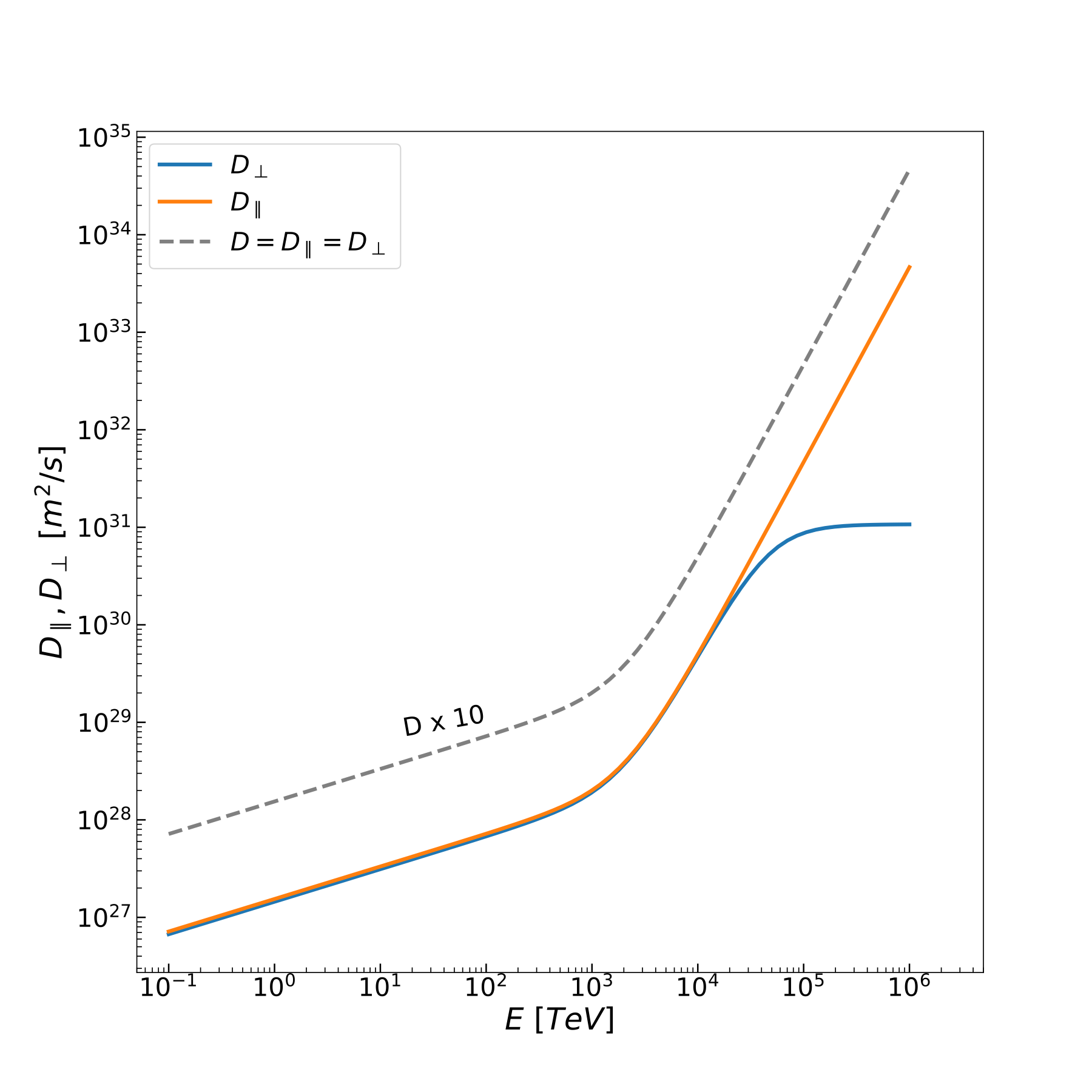}
\caption{Energy dependence of the diffusion coefficients $D_\parallel$ and $D_\perp$. Model I: $D_\parallel = D_\perp$, Model III: $D_{\parallel} \ne D_{\perp}$.}
\label{fig:D}

\end{minipage}
\end{figure*}
and perpendicular -- as well as the effect of CR anisotropic propagation, three models (I, II and III) were constructed, each with distinct propagation conditions, magnetic field configurations, and forms of the diffusion tensor. The primary objective was to analyze how variations in the energy dependence of the diffusion tensor coefficients and propagation anisotropy influence the CR spectra and the spatial distribution of CR densities. \par

\textbf{A. Model I: Isotropic diffusion with identical energy dependencies}\par

In this model, a thin Galactic disk surrounded by a halo was considered, as illustrated in Fig.\ref{fig:galaxy}. The red color schematically depicts the Galactic halo with dimensions a=17 kpc and b=6 kpc. The green color indicates the thin disk with a thickness of h=0.4 kpc. The color scale represents the magnitude of the regular magnetic field component, ranging from approximately 1 $\mu G$ in the Galactic halo (dark purple) to 6 $\mu G$ in the spiral arms within the disk. The magnetic field comprises a regular component based on the Unger24 model\cite{Unger_2024}, as well as a turbulent component of constant power. The source distribution follows the model from~\cite{yusifov2004galacticdistributionluminosityfunction}. It is important to note that a source distribution strictly confined to spiral arms would enhance the observed effects. Therefore, a simplified source distribution~\cite{yusifov2004galacticdistributionluminosityfunction} was employed to clearly demonstrate the influence of the Galactic magnetic field. The diffusion tensor components along and across the magnetic field ($D_\parallel$ and $D_\perp$ ) were assumed equal and followed the same energy dependence with a spectral break around $10^6$ GeV (the diffusion coefficient D is marked with a gray dashed line in Fig.~\ref{fig:D}), corresponding to an almost isotropic scenario.

As a result of solving Eq. (\ref{eq:difusion_eq}) for the Galactic volume defined in model I, a “universal” knee appears in the CR spectrum at 4 PeV. The spectral indices before and after the break are determined solely by the behavior of the diffusion coefficients, and the difference between them ($\Delta_\gamma$) remains uniform throughout the Galaxy. This results in a sharper spectral transition than observed experimentally. The spatial distribution of CR density fully mirrors the source distribution.\par

\begin{figure*}[htbp]
\centering

\begin{minipage}{0.45\textwidth}
\centering

\includegraphics[width=\textwidth]{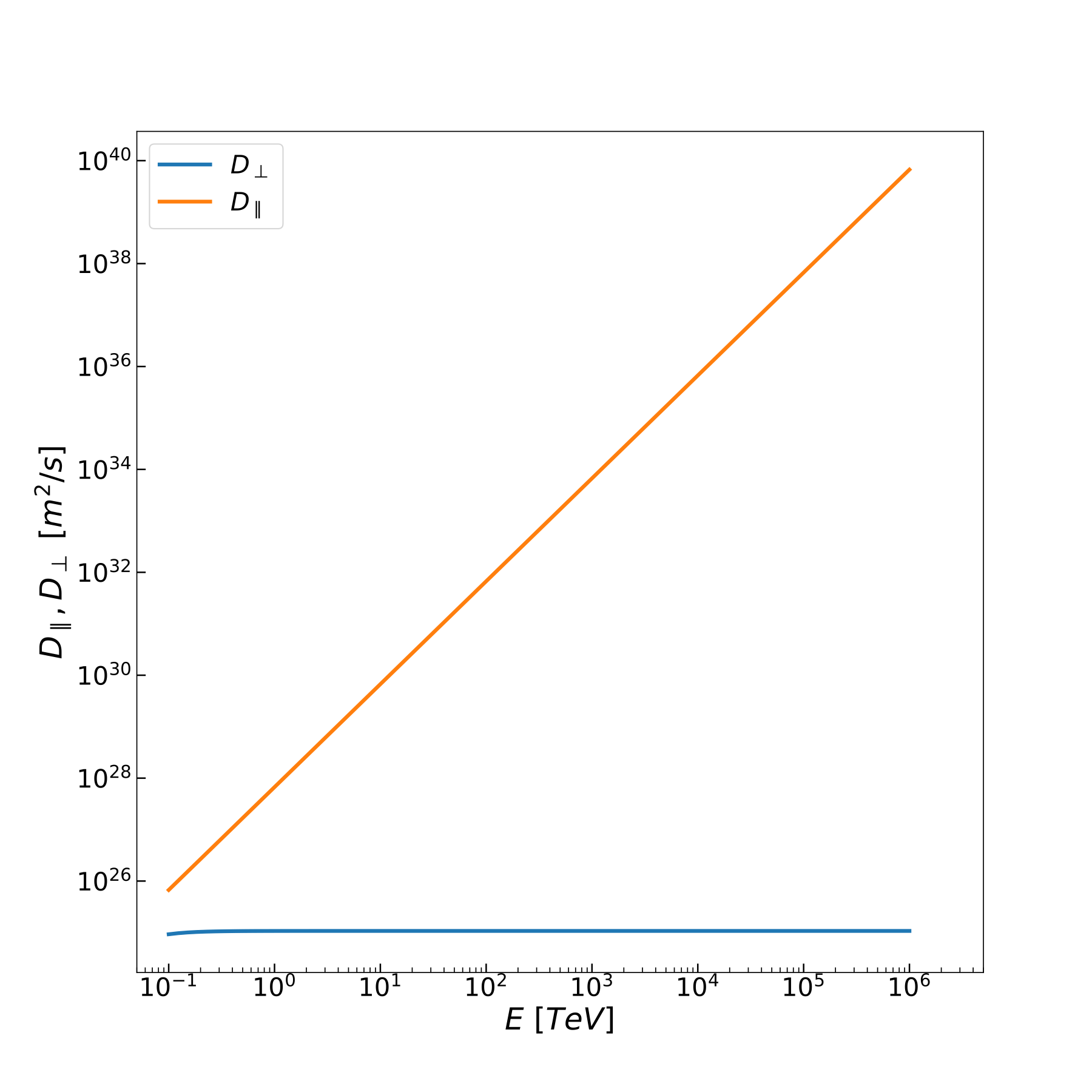}
\caption{Energy dependence of the diffusion coefficients 
$D_{\parallel}$ and $D_{\perp}$ . Model II: no spectral break in the energy dependence.}
\label{fig:2D}

\end{minipage}%
\hfill
\begin{minipage}{0.45\textwidth}
\centering



\includegraphics[width=\textwidth]{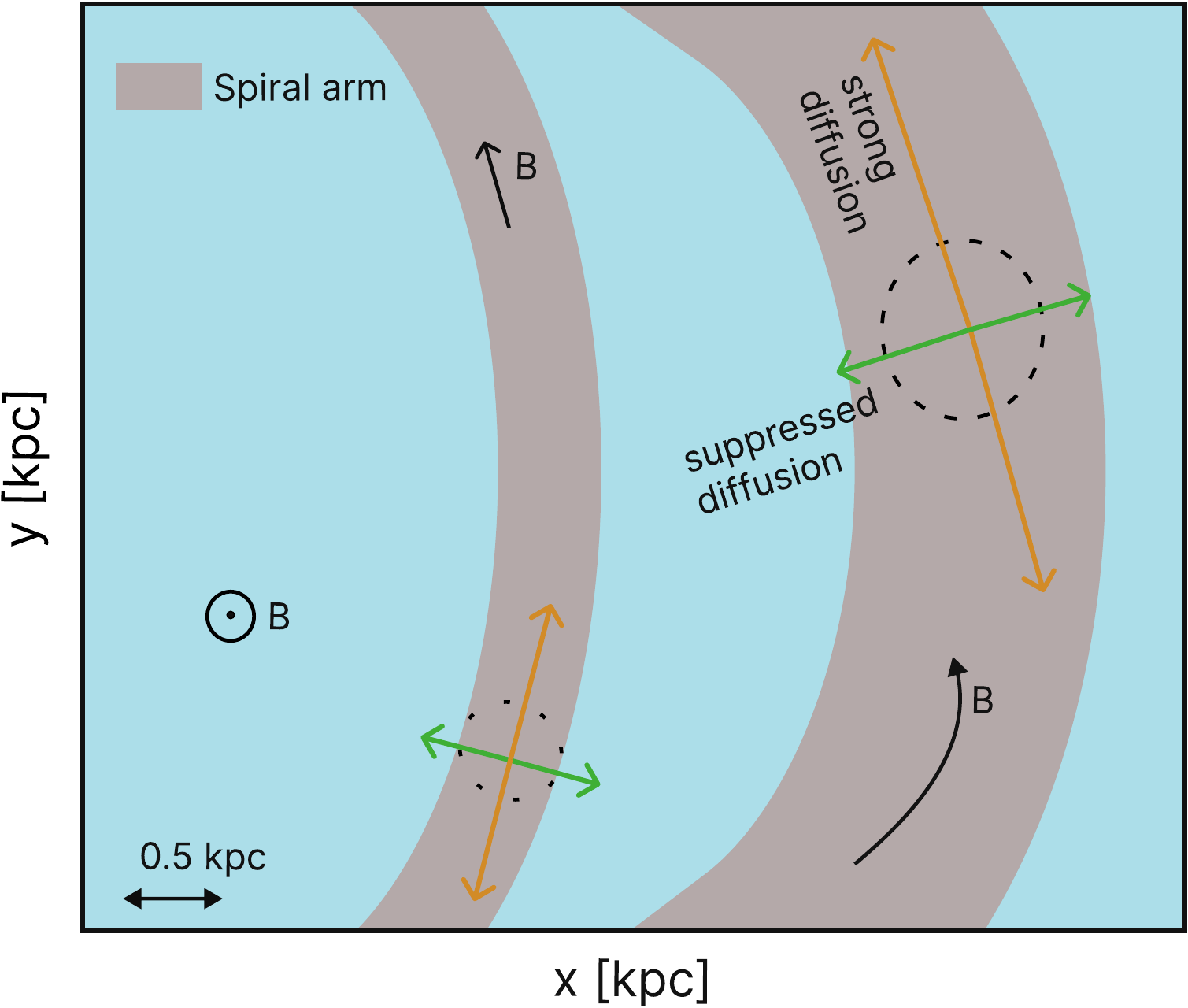}
\caption{Scheme of a modeled region at a constant z slice, showing the preferred diffusion direction.}
\label{fig:spur}

\end{minipage}
\end{figure*}

\textbf{B. Model II: Anisotropic diffusion with different energy dependence of $D_\parallel$ and $D_\perp$}\par

In this model, the geometry of the Galaxy and the source distribution remain the same as in model I (see Fig. \ref{fig:galaxy}). However, a distinct energy dependence of the components of the diffusion tensor was introduced (see Fig. \ref{fig:2D}), with $D_\parallel$ significantly exceeding $D_\perp$ throughout the entire energy range. This implies that particles predominantly diffuse along the magnetic field lines. The slopes of the diffusion tensor components remain unchanged across the considered energy range.

To isolate the effect of anisotropy, the magnetic field component perpendicular to the disk ($B_z$) was artificially set to zero in this model. This highlights the impact of transverse diffusion, since even a small perpendicular magnetic field contribution may lead to significant particle leakage when $D_\parallel \gg D_\perp$. From a physical perspective, the difference between the longitudinal and transverse diffusion coefficients results in a nonisotropic net displacement direction for the particles. Figure \ref{fig:spur} illustrates a model region within the Galactic disk that includes two spiral arms. The regular magnetic field within the arms is uniformly directed and corresponds to a specific predefined orientation. Outside the arms, the magnetic field is predominantly oriented perpendicular to the disk. The regions from which diffusion is observed are indicated by dashed circles. Since the field is aligned along the Galactic arm and the transverse diffusion coefficient is suppressed, the redistribution of CR concentration will primarily occur along the spiral arm (orange vectors). Diffusion perpendicular to the arms is minimal (green vector). As a result, the CR concentration outside the arms is strongly suppressed compared to that within the arms, and this effect becomes more pronounced with increasing energy, since $D_\parallel$ increases rapidly with energy, while $D_\perp$ remains constant. Conversely, the relative CR concentration inside the arms at high energies increases (with the growth rate determined by the degree of propagation anisotropy), and at high energies, more particles of a given energy will be observed than are injected by the source at the same energy--i.e., the spectrum becomes harder within the arms and softer in the regions with strong outflow.
\begin{figure*}[htbp]
\centering

\begin{minipage}{0.45\textwidth}
\centering


\includegraphics[width=\textwidth]{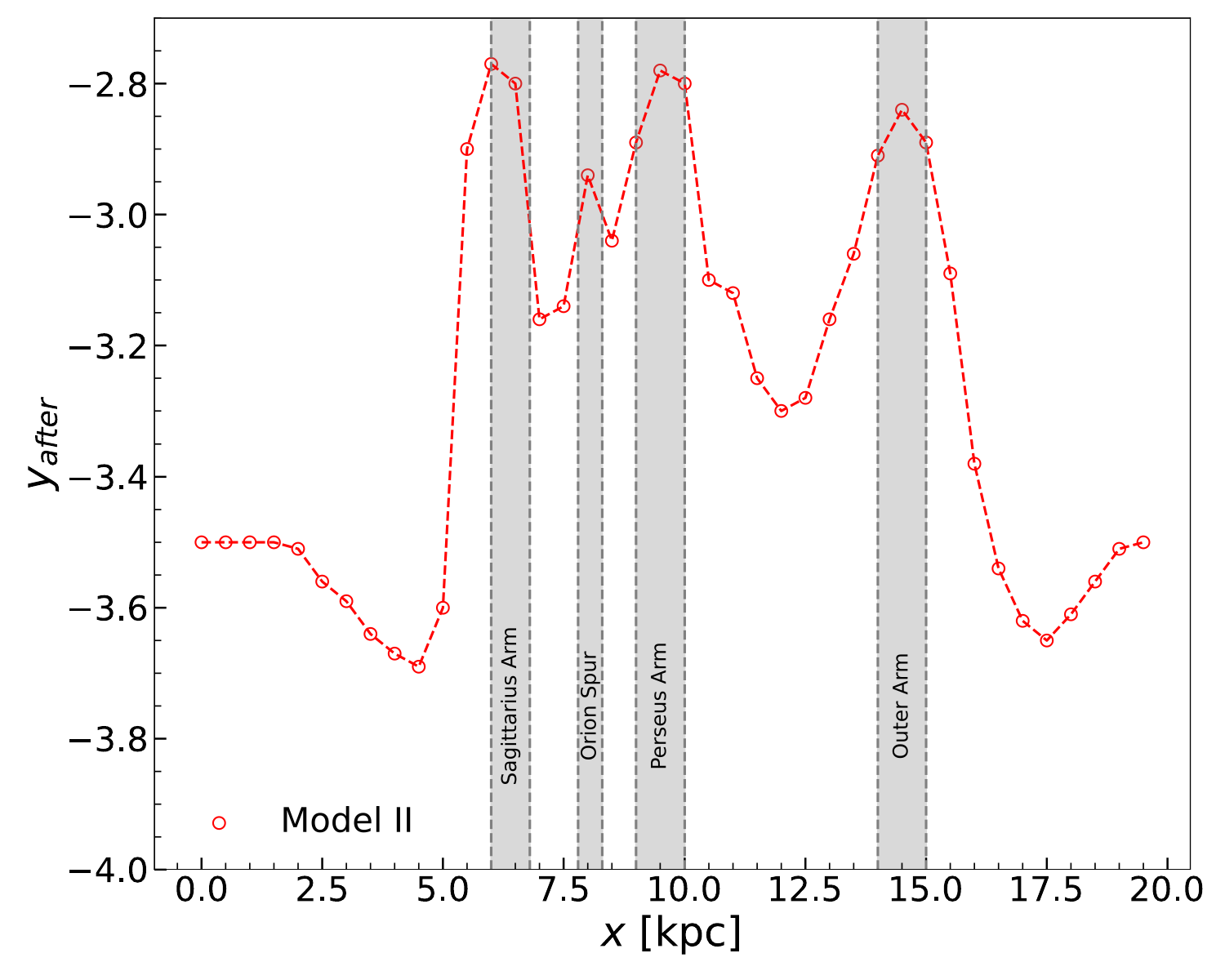}
\caption{Spectral index hardening in model II as a function of distance from the Galactic center, modified due to anisotropic CR propagation. Spiral arm regions of the Galaxy are shaded in gray.}
\label{fig:g_break}
\end{minipage}%
\hfill
\begin{minipage}{0.45\textwidth}
\centering


\includegraphics[width=\textwidth]{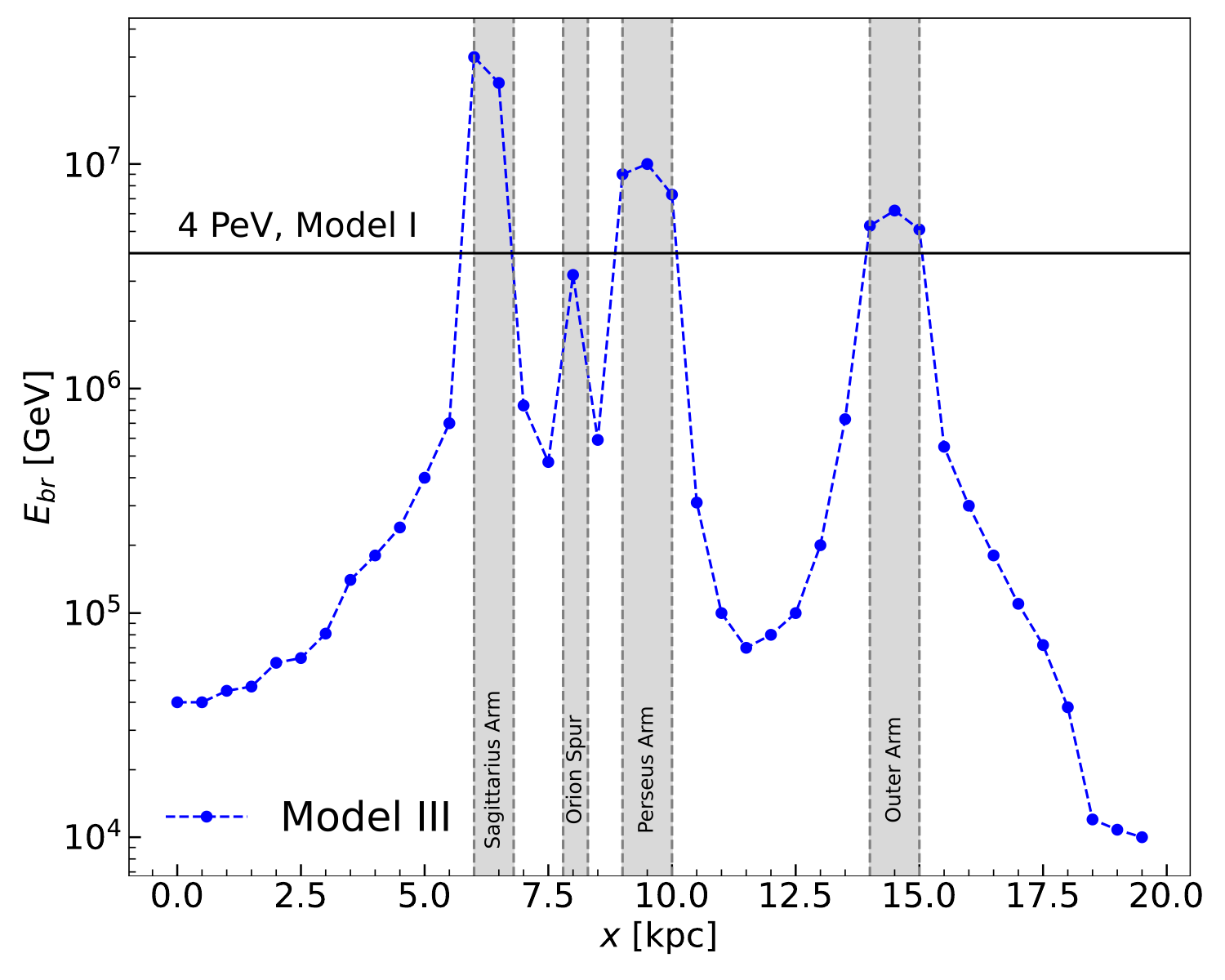}
\caption{Position of the CR spectral knee as a function of distance from the center. The blue dashed line indicates the position of the break in model III; the solid black horizontal line shows the "universal" knee in model I.}
\label{fig:E_break}

\end{minipage}

\end{figure*}

\par
In model II, a clear spectral knee is absent. However, Fig. \ref{fig:g_break} shows that the spectral index varies: it becomes harder in regions where particles accumulate due to diffusion along the field, and softer in zones of active particle "wash-out." Thus, the spatial distribution of CRs at high energies deviates from the source distribution and is instead governed by the geometry of the magnetic field. The anisotropy of the distribution increases with energy.

\textbf{C. Model III: Complex case with varying turbulence.}\par
Model III combines the effects considered in models I and II. The structure of the Galaxy and the source distribution follow those of model I (see Fig. \ref{fig:galaxy}). The energy dependence of the diffusion tensor components is illustrated in Fig. \ref{fig:D}, where the longitudinal diffusion coefficient is shown in orange and the transverse one in blue. The strength of the turbulent magnetic field varies throughout the Galaxy and is defined by the equation:
\begin{equation}
\eta(x,y,z) =\eta_0 \cdot \exp\left(-\frac{R}{r_0}\right)\cdot \exp\left(-\frac{z}{z_0}\right),
\label{eq:turb}
\end{equation}
where $\eta_0 = B_{turb}/B_{reg}$, $r_0 = 5$ kpc, and $z_0 = 1$ kpc.
The turbulent field decreases exponentially with distance from the Galactic center both radially and vertically (along the z axis).

In the course of solving Eq. (\ref{eq:difusion_eq}), CR concentrations were computed in the modeled 3D volume for each energy, and the spectral break position was determined. As shown in Fig. \ref{fig:E_break}, the position of the CR spectral break shifts: in regions with enhanced magnetic field strength (both regular and turbulent), it moves toward higher energies, whereas in zones with weaker fields, it shifts to lower energies. The variation range of the break is approximately from $10^4$ to $10^7$ GeV. The spectral slope also varies: it becomes harder in the Galactic arms, where the field is stronger, and softer outside of them (as discussed in model II). Here, the energy location of the spectral break in model I is indicated by the solid black line.

Thus, the energy dependence of the diffusion tensor components leads to the formation of a spectral break in the CR spectrum, with its exact position determined by the structure of the magnetic field. The inhomogeneous distribution of the turbulent magnetic field throughout the Galaxy enhances this effect, as variations in the turbulence affect the correlation length, which implicitly enters the expression for $D$. The spectral slope before and after the break is determined by the combined effect of the change in the power-law behavior of the diffusion coefficients and the anisotropic propagation effects. As in Model II, the spatial distribution of CR concentrations at high energies deviates from the source distribution and is governed by the geometry of the magnetic field.

By varying the source distribution parameters the characteristics of the regular and turbulent magnetic fields, as well as the turbulence correlation length, it is possible to reproduce the observed CR spectrum at Earth without invoking cutoff energy in the sources and to make predictions for other regions of the Galaxy.

\section{\label{sec:all_elements}Spectrum of All Elements}

Our calculations demonstrate that CR transport plays a key role in shaping the energy spectrum, including the formation of the spectral break known as the CR knee. To explore this, we calculate element-specific CR spectra, as well as the total spectrum across the energy range from above the solar modulation cutoff up to approximately $10^5$ -- $10^6$ TeV --- the practical limit for diffusion-based modeling. In these calculations, we assume a simple power-law source spectrum with index varying as $\gamma_{source} \in [-2.4; -2.0]$, without imposing an explicit acceleration cutoff, in order to isolate the effects of transport.

Elemental CR spectra in the knee region are typically described by a broken power-law function, where the spectral index $\gamma_1$ before the break transitions to $\gamma_2$ after the break.

In the study~\cite{borisov2025modulationgalacticcosmicray}, the authors demonstrated modulation of the CR proton spectrum around 4 PeV and described the leakage of CR concentrations into the Galactic halo. The spectrum of all CR elements can be obtained using a similar approach and described by the function:

\begin{eqnarray}
F_s(\gamma_1, \gamma_2, N_0, E_0, E) = 
N_0 \left(\frac{E}{E_0}\right)^{-\frac{\gamma_1 + \gamma_2}{2}}&&\times\nonumber\\
\left[
\frac{\left(\frac{E_0}{E}\right)^{\frac{s}{2}} + \left(\frac{E}{E_0}\right)^{\frac{s}{2}}}{2}
\right]^{\frac{\gamma_1 - \gamma_2}{s}}.
\label{eq:dist}
\end{eqnarray}

\begin{figure}[h]
\includegraphics[width=0.5\textwidth]{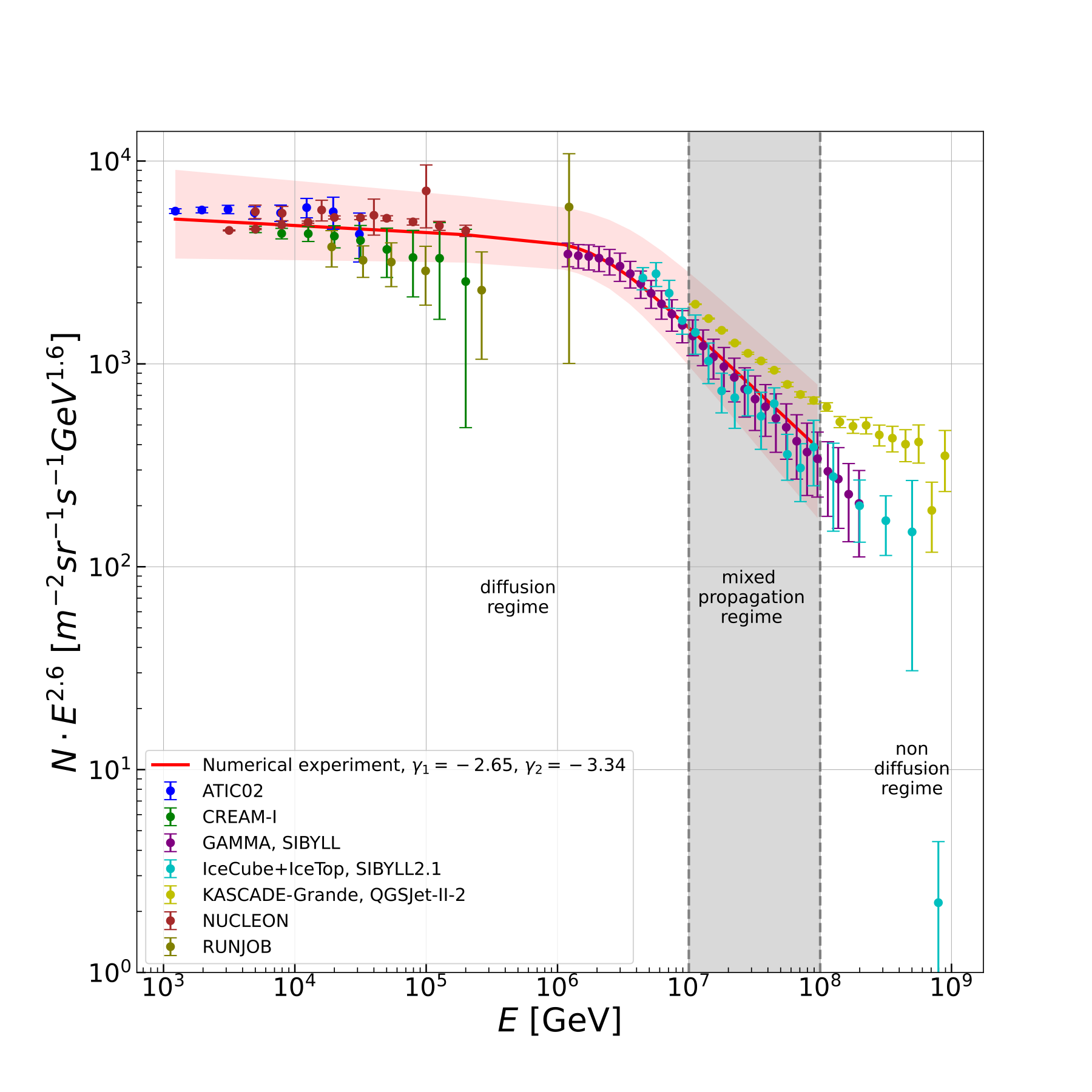}
\caption{\label{fig:p_spec} Modeled energy spectrum of CR protons compared with experimental data from  ATIC~\cite{Panov2007}, CREAM~\cite{Ahn2010}, GAMMA~\cite{Garyaka2008}, IceTop~\cite{Aartsen_2019}, KASCADE-Grande~\cite{Apel2012}, NUCLEON~\cite{Grebenyuk2019} and RUNJOB~\cite{Apanasenko2001}. The gray dashed line marks the approximate boundary of the applicability of the diffusion approach.}
\end{figure}

Here, $\gamma_1$ and $\gamma_2$ represent the spectral indices before and after the break, respectively, $E_0$ is the break position, $N_0$ is the number density at the break point, and $s$ is the smoothing parameter. For each element group, the calculated concentrations were approximated by function \ref{eq:dist} with minimization over five free parameters: $(\gamma_1, \gamma_2, N_0, E_0, E)$. Figure \ref{fig:p_spec} presents the modeled CR proton spectrum, where the best-fit line is shown in red with an error range of $\pm \sigma$.

The primary objective was to adjust the input model parameters to achieve a reasonable description of the break position, spectral slopes, and observed flux. Due to inconsistencies among various experiments and their differing statistical precision, we did not aim to fit a specific experiment but rather sought to capture the key features of the spectrum.

To describe the spectrum of all elements, we use the conventional approach of grouping elements into four categories: protons, helium, light, intermediate, and heavy elements. For CR concentrations, we use data from ATIC~\cite{Panov2007}, CREAM~\cite{Ahn2010}, GAMMA~\cite{Garyaka2008}, IceTop~\cite{Aartsen_2019}, KASCADE-Grande~\cite{Apel2012}, NUCLEON~\cite{Grebenyuk2019} and RUNJOB~\cite{Apanasenko2001}  for protons and individual element groups, along with data from Auger~\cite{novotný2025energyspectrummasscomposition}, HAWC~\cite{disciascio2022measurementenergyspectrumelemental}, and TUNKA~\cite{Budnev_2020}.

\begin{figure*}
\includegraphics[width=0.95\textwidth]{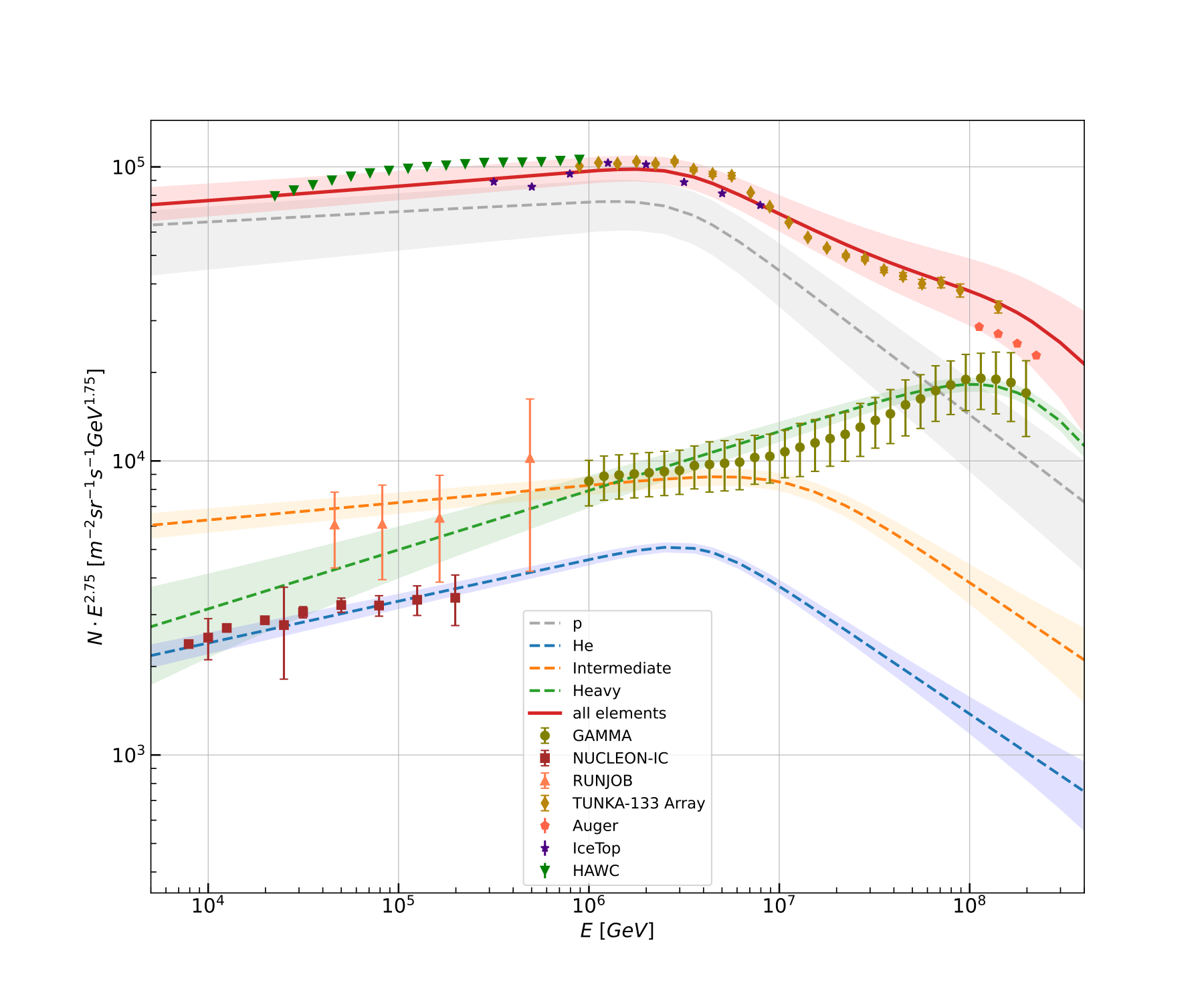}
\caption{\label{fig:all_spectra}Energy spectrum of all elements.}
\end{figure*}

It is well known that data for individual element groups are statistically much less reliable compared to data for all nuclei. An exception is proton statistics, which remain satisfactory due to their high abundance. Figure \ref{fig:all_spectra} presents the spectrum of all elements along with component spectra for individual nucleus groups.

The CR concentrations within a given energy range are calculated at each node of the simulated Galactic volume, allowing us to evaluate CR energy spectra in different regions of the Milky Way, such as outside spiral arms, in the bulge, or in the halo at some distance from the disk. The strong dependence of the diffusion tensor coefficients and the position of the critical point on the magnetic field magnitude results in a dependence of the CR knee energy on the magnetic field strength and, consequently, on the spatial location within our Galaxy.

\begin{figure}[h]
\includegraphics[width=0.45\textwidth]{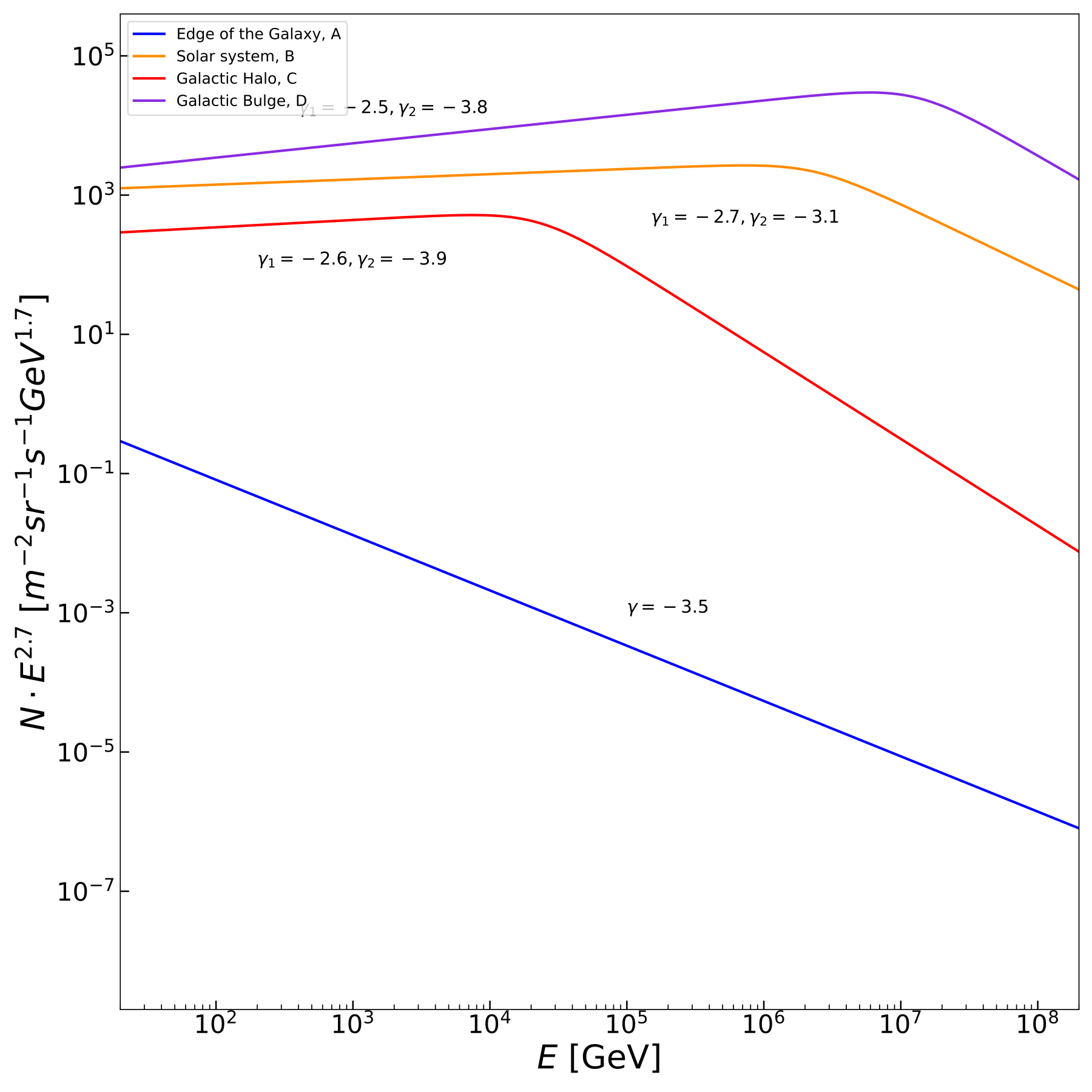}
\caption{\label{fig:galaxy_spec} Energy spectra of CR p for different spatial locations within the Milky Way.}
\end{figure}

Figure \ref{fig:galaxy_spec} presents the spectra of CR protons and the corresponding spectral slopes before and after the knee for points A, B, C, and D, shown in Fig. \ref{fig:galaxy}. The positions were chosen to highlight differences in the resulting spectrum, as its behavior varies only slightly between closely spaced grid nodes, according to the procedure described in Sec. \ref{sec:anis_features}. Four representative locations are shown: two in the Galactic disk (points B and D) and two in the halo (points A and C). Their Cartesian coordinates (in kpc) relative to the Galactic center are as follows: B (-8.2,0.0,0.2), D (6.4,0.0,0.0), A (-18.0,-18.0,-5.8), and C (5.4,0.0,-2.0). These points are selected to illustrate the spatial variation of the CR knee (see Sec. \ref{sec:anis_features} for details). In the model, spectra were computed for all nuclear groups at every node of the finite-difference grid. The spatial resolution was adapted to ensure accurate modeling of diffusion at the relevant energy scales: 0.1 kpc for 1 TeV--10 PeV, 0.5 kpc for 10 PeV--100 PeV, and 5 kpc for the range of limited applicability of the diffusion approximation (100 PeV--1000 PeV), where mixed transport regimes may dominate (see Sec. \ref{sec:Method} for details). It is important to note the limited validity of the diffusion approach in this highest energy range. The modeled spatial volume extends from $-20$ to $20$ kpc in both the $x$ and $y$ directions and from $-6$ to $6$ kpc in $z$, as illustrated in Fig. \ref{fig:galaxy}. It is evident that the break position is determined by the critical point, and in regions with significantly different diffusion tensor components, it shifts to higher energies (point D), whereas in regions located farther from the disk (point C), the break position moves toward lower energies. As the distance to the Galactic boundary increases, the difference $\Delta \gamma$ decreases and eventually reaches zero, which results from the diminishing impact of transport as the field isotropizes.

\section{\label{sec:results}RESULTS}

At present, the only method for obtaining information about the structure of the CR energy spectrum in regions of the Galaxy other than the Solar System's location is gamma-ray sky survey observations. As CRs propagate through the Galaxy, they interact with diffuse clouds and interstellar dust, producing $\pi^+$, $\pi^-$, and $\pi^0$ mesons~\cite{Ackermann_2012}. Gamma-ray photons generated in meson decays, being neutral particles, do not participate in diffusive propagation and carry information about the energy spectrum structure of primary CR elements in their production region.

Currently, the Fermi-LAT and LHAASO experiments have provided sky survey data in the energy range up to 1 PeV. Both experiments have examined the same sky region within the angular range $ 15^{\circ} < l < 125^{\circ}, |b| < 5 ^{\circ}$ (the inner Galaxy) and $ 125^{\circ} < l < 235^{\circ} , |b| < 5 ^{\circ}$ (the outer Galaxy) in the Galactic coordinate system, enabling their use for analyzing primary CR spectra.

The CR spectrum reconstructed from gamma-ray observations shows poor agreement with the "universal" CR knee model, as demonstrated in~\cite{prevotat2024energydependencekneecosmic}. To describe the available experimental data, it would be necessary to assume that the spectral break, known as the CR " knee" occurs at an energy 1--2 orders of magnitude lower than its observed position, which is certainly not the case near Earth.

Our results indicate that the CR knee is a spatially dependent feature: its position, along with the spectral indices $\gamma_1$ and $\gamma_2$, varies across different regions of the Galaxy (See Sec. \ref{sec:anis_features} for details). Our calculations show that accounting for the energy dependence of the diffusion tensor components and the anisotropic propagation of cosmic rays allows one to describe the spectra of high-energy diffuse gamma rays, explaining their behavior by the emerging spatial dependence of the CR knee. This conclusion is based on computed primary spectra for protons (p), helium (He), medium-mass nuclei ($Z = 6$–$9$), and heavy nuclei ($Z \geq 10$), obtained within the framework of anisotropic diffusive propagation. The ultraheavy element group was not considered due to its negligible contribution to the resulting gamma-ray spectrum. To determine the energy spectra of gamma-ray photons, we utilized the aafrag package~\cite{Koldobskiy_2021}, based on the QGSJET-II nuclear interaction model~\cite{Ostapchenko2006}, for calculating cross sections of interactions between CR nuclei and protons with distributed matter in the Galaxy.

For the matter distribution, we adopted the simplest parametrization, as in the case of source distribution~\cite{Lipari_2018}. The distribution of neutral atomic (H), molecular (H$_2$), and ionized hydrogen was considered, with matter primarily concentrated in the Galactic disk, exponentially decreasing toward the edges and the Galactic halo. In cylindrical coordinates, the matter distribution is described by the following equation:

\begin{equation}
n(r,z) = n_0(r)\exp\left[-\frac{z^2}{2\sigma_z^2(r)}\right],
\end{equation}

where $\sigma_z^2(r)$ represents the average density in the disk center. Detailed parameter values for each Galactic region are provided in~\cite{Lipari_2018}.

In~\cite{prevotat2024energydependencekneecosmic}, it was demonstrated, that in CR interactions with interstellar gas at energies above $\geq 10$ TeV, the production channels of $\pi^+$ and $\pi^-$ mesons are suppressed by approximately an order of magnitude compared to $\pi^0$ meson production. Thus, in a first approximation, these channels can be neglected in spectrum calculations. In this case, the total gamma-ray flux is given by the equation

\begin{equation}
\begin{split}
I_{\gamma}(E,l,b) &= \frac{c}{4\pi}\Sigma_{A,A'} \int_{0}^{\infty} n_{gas}^A(x)e^{-\tau (s,E)} ds \\
&\times \int_{E}^{\infty}\frac{d\sigma^{A',A \rightarrow X_{\gamma}(E',E)}}{dE}
\frac{dN_{CR}^{A'}}{dVdE'}dE'
\end{split}
\end{equation}

where $x={s,l,b}$ represents a three-dimensional spherical coordinate system ($s$ being the distance along the line of sight in the direction $(l,b)$ in Galactic coordinates), and $\frac{d\sigma^{T,A \rightarrow X_{\gamma}}(E', E)}{dE}$ is the differential cross section for photon production with energy $E$ in the interaction of a primary nucleus $A'$ of energy $E'$ with a target nucleus $A$ of density $n_{gas}^A$. The optical depth $\tau$ accounts for photon absorption due to pair production on cosmic microwave background (CMB) photons, extragalactic background light, dust, and starlight. Notably, the absorption peak occurs around 2.2 PeV, as shown in~\cite{Vernetto_2016}.

\begin{figure*}[htbp]
\centering
\begin{minipage}{0.45\textwidth}
\centering
\includegraphics[width=\textwidth]{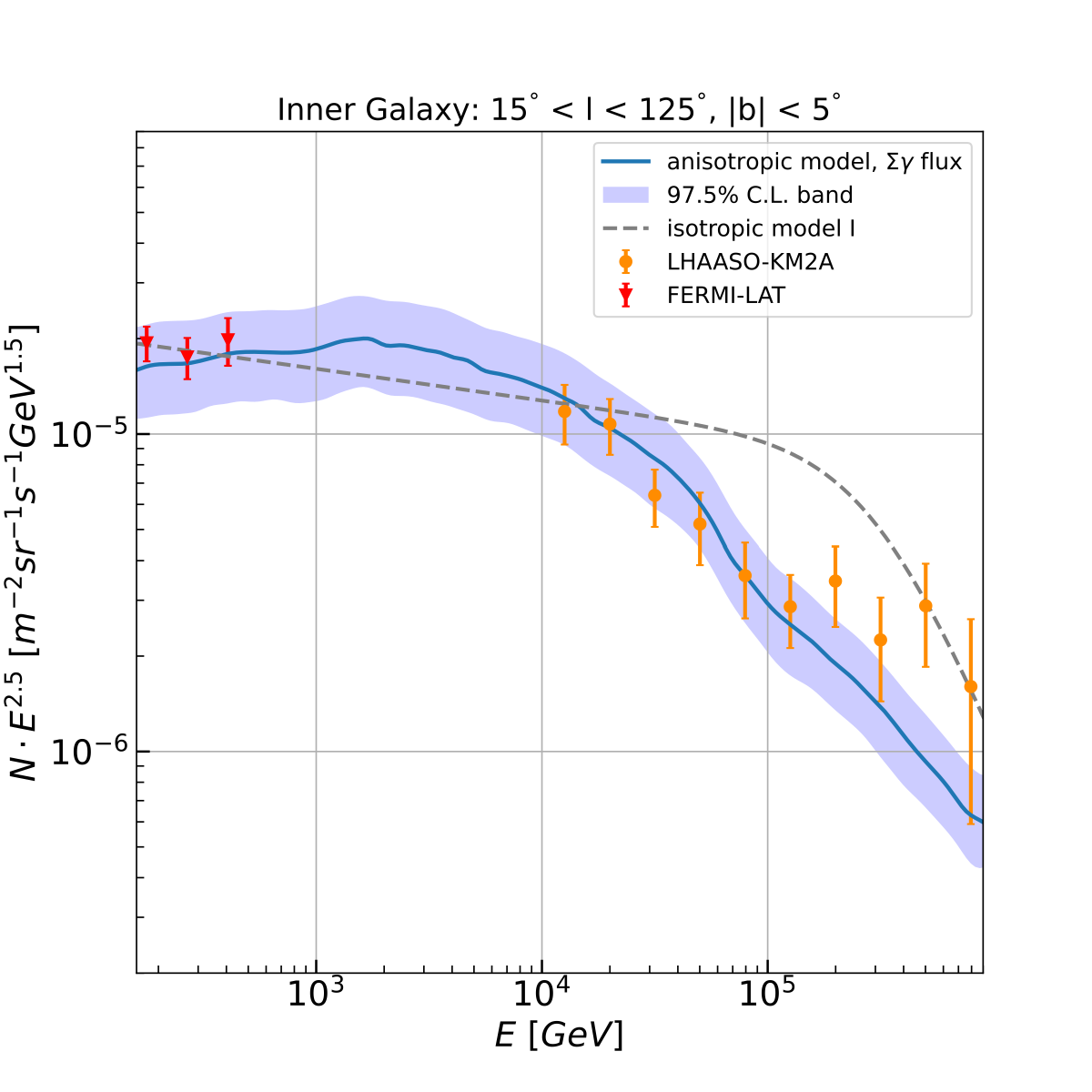}
\caption{Gamma-ray spectrum in the inner Galactic region for LHAASO ~\cite{Cao_2023} and Fermi-LAT ~\cite{Zhang_2023}.}
\label{fig:gamma_inner1}
\end{minipage}%
\hfill
\begin{minipage}{0.45\textwidth}
\centering
\includegraphics[width=\textwidth]{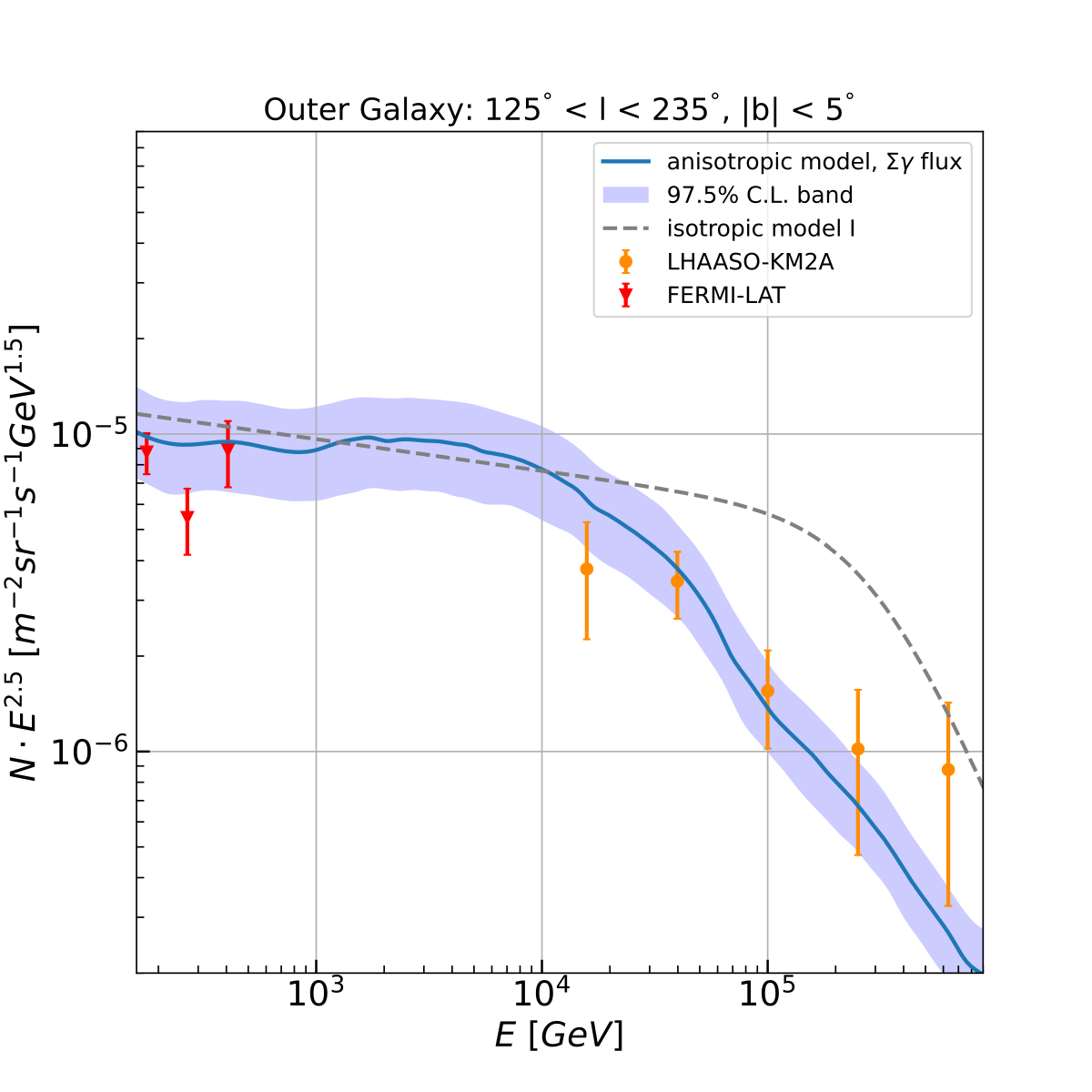}
\caption{Gamma-ray spectrum in the outer Galactic region for LHAASO ~\cite{Cao_2023} and Fermi-LAT ~\cite{Zhang_2023}. }
\label{fig:gamma_outer}
\end{minipage}
\end{figure*}

Figures \ref{fig:gamma_inner1} and \ref{fig:gamma_outer} present the modeled gamma-ray spectrum (shown as a blue line, computed within the anisotropic diffusion approach) along with the 99.7$\%$ confidence interval for the inner and outer Galaxy regions. These are compared against experimental observations from Fermi-LAT and LHAASO for the corresponding regions. The confidence interval reflects the numerical uncertainty due to variations in the source spectral index within the range $\gamma_{source} \subseteq [-2.4; -2.0]$.The modeled gamma-ray spectrum is consistent with the experimental data, remaining largely at the same level though it exhibits a slightly steeper decline near the PeV range. In our model, this steepening arises from accounting for gamma-ray attenuation on CMB during propagation.

The gray dashed line shows the integrated diffuse gamma-ray spectrum obtained using model I (described in Sec. \ref{sec:anis_features}). A similar result is reported in study ~\cite{prevotat2024energydependencekneecosmic}. Transitioning to an anisotropic cosmic-ray diffusion approach with a spatially dependent knee energy allows for a more accurate reproduction of the observed gamma-ray spectrum. As shown in Sec. \ref{sec:anis_features} and illustrated in Fig. \ref{fig:galaxy_spec}, the cosmic-ray knee emerges at lower energies in interarm regions and toward the edges of the Galaxy, leading to a substantial modification of the resulting gamma-ray spectrum.

\section{\label{sec:conclusion}CONCLUSION}

The numerical calculations of CR proton and nuclei spectra presented in this work are based on the stationary anisotropic diffusion equation, with diffusion tensor components derived using a trajectory-based method. These results reveal a spatial dependence of the spectral break -- commonly known as the cosmic-ray knee -- across the Galaxy. The constructed model reproduces the CR knee without requiring a high-energy cutoff in the source spectrum. Instead, the knee arises naturally from the anisotropic nature of CR propagation and the energy dependence of the diffusion tensor components.

To investigate this effect, three models were developed, each incorporating different assumptions about the size of the simulated Galactic volume, the magnetic field structure, and the diffusion tensor. It is shown that the CR knee can be explained by two key factors: (i) the position of the spectral break is governed by the spatial dependence of the diffusion tensor components, which are themselves determined by the magnetic field model; and (ii) the resulting slope of the CR spectrum is shaped by anisotropic diffusion, which arises from the directional dependence of CR propagation.

This spatially varying knee leads to modified gamma-ray spectra throughout the Galaxy. The predicted spectra, when compared with gamma-ray observations from LHAASO and Fermi-LAT, demonstrate improved agreement over standard isotropic diffusion models. This suggests that incorporating anisotropic CR transport with spatially dependent spectral features provides a more accurate framework for interpreting high-energy astrophysical data.

\appendix

\section{\label{app:subsecA}MODEL PARAMETERS}

There are two main approaches~\cite{kuhlen2023},~\cite{Kudryashov2023} to the numerical integration of test-particle trajectories for computing the components of the diffusion tensor $ D_{ij}({B(\mathbf{r})},E)$, depending on the magnetic field model and turbulence configuration. These methods result in different energy dependencies of the parallel and perpendicular diffusion coefficients, particularly below the transition region where the diffusion regime changes. In this work, we adopt the integration approach proposed in~\cite{Kudryashov2023} and apply it to compute the diffusion coefficients for specified values of the regular and turbulent magnetic fields, the turbulence correlation length $\lambda_c$, and the spectral index of turbulence. For a given set of parameters we compute the energy dependence of each component of the diffusion tensor $ D_{ij}({B(\mathbf{r})},E)$ using test-particle trajectory integration. These results are also used to determine characteristic energy step sizes for the finite-difference scheme. The parameterization is then performed for the given set of parameters. However, despite the parametrization of the energy dependence, the full diffusion tensor $ D_{ij}({B(\mathbf{r})},E)$ is recomputed in each grid cell of the simulation domain. This is done by rescaling the precomputed energy dependencies according to the local values of the regular and turbulent magnetic field components in each node. This approach ensures that anisotropy and spatial variation of diffusion are preserved across the grid. Furthermore, the integration procedure can be repeated for different choices of the correlation length or turbulence spectra, including non-Kolmogorov cases, to produce alternative scalings applicable to different astrophysical conditions.

The diffusion coefficients used in this work are given by the expression (\ref{eq:cases}), which approximate the results of numerical integration of Eqs. (\ref{eq:5}) and (\ref{eq:6}) for a turbulence correlation length $\lambda_c = 100$ pc and a Kolmogorov power spectrum as defined by Eq. (\ref{eq:kolmog}. These energy-dependent forms are recalculated in each cell of the simulation grid according to the local values of the regular and turbulent magnetic field components. Since the exact value of the correlation length $\lambda_c$ in the interstellar medium is not well constrained observationally, it remains a free parameter. Varying $\lambda_c$ in combination with fitting to experimental data can provide a more accurate estimate of its value. This procedure will be addressed in future work.

\begin{equation*}
 \begin{array}{l}
   \lambda_1(E) = 0.2 E^{a} + 4 E^2, 
   \\
   \lambda_\parallel(E) = 3 E^{a} + 3 E^2,
   \\
    \lambda_\perp(E) = 0.09E^{a}\exp(-\frac{E}{x_0}) + \\0.25(1-\exp(-\frac{E}{x_0})), 
    \\
    D_\parallel ({B(\mathbf{r})},E)  = b \left(\lambda_1(\frac{E}{{B(\mathbf{r})}})+\lambda_\parallel(\frac{E}{{B(\mathbf{r})}})\right),
    \\
     D_\perp ({B(\mathbf{r})},E)  = b \times \\ \left(\lambda_1(\frac{E}{{B(\mathbf{r})}})^{-1} +\lambda_\parallel(\frac{E}{{B(\mathbf{r})}})^{-1} \right)^{-1}.
     \label{eq:cases}
 \end{array}
\end{equation*}
Here, \(E\) is the particle energy, \(x_0 = 0.3\) is a normalization constant, and \(b = 1.67 \cdot 10^{23}\) [m\(^2\)/c] ensures correct dimensionality. The components are computed in the magnetic field-aligned coordinate system, assuming equal strengths of regular and turbulent components, and are then rotated into Cartesian coordinates for use in the finite-difference scheme. Two options for the exponent \(a\) are considered. The case \(a = 1/3\) leads to source spectral indices in the range \([-2.4; -1.8]\), consistent with shock acceleration models. In contrast, for \(a \in [-1.2; 0]\), which reflects variations in field strength, the resulting steeper indices are less consistent with pure diffusive models. We suggest this tension could be resolved by including preacceleration and energy loss processes in future refinements.

\begin{acknowledgments}
The authors express their gratitude to Professor Panov A. D. for valuable discussions during the preparation of the article. The research was supported by RSF (Project No. 25-22-00246).
\end{acknowledgments}

\section*{DATA AVAILABILITY}
The experimental data used in this work are publicly available from the sources cited in Refs. ~\cite{Cao_2023} and ~\cite{Zhang_2023} (gamma-ray data). For a more detailed description of the mathematical framework employed in this study,
see Refs. ~\cite{kuhlen2023} and ~\cite{Kudryashov2023}. Data related to cosmic-ray nuclei were obtained from Refs. ~\cite{Panov2007, Ahn2010, Garyaka2008, Aartsen_2019, Apel2012, Grebenyuk2019, Apanasenko2001, novotný2025energyspectrummasscomposition, disciascio2022measurementenergyspectrumelemental, Budnev_2020}. The code developed
for data processing and analysis is available from the corresponding author upon reasonable request.


\begin{thebibliography}{99}

\bibitem{H_randel_2003}
J.~R. Hörandel.
\newblock On the knee in the energy spectrum of cosmic rays.
\newblock \href{https://doi.org/10.1016/S0927-6505(02)00198-6}{{\em Astroparticle Physics}, 19(2):193–220, 2003.}

\bibitem{Kachelrie__2019}
M.~Kachelrieß and D.~V. Semikoz.
\newblock Cosmic ray models.
\newblock \href{https://doi.org/10.1016/j.ppnp.2019.07.002}{{\em Progress in Particle and Nuclear Physics}, 109:103710, 2019.}

\bibitem{Zirakashvili2021}
V.~N. Zirakashvili and V.~S. Ptuskin.
\newblock Cosmic Ray Acceleration in Supernova Remnants with Nonuniform Density Distribution.
\newblock \href{https://doi.org/10.3103/S1062873821040407}{{\em Bulletin of the Russian Academy of Sciences: Physics}, 85(4):404–407, 2021.}

\bibitem{Lagutin2001}
A.~A. Lagutin and V.~V. Uchaikin.
\newblock Fractional diffusion of cosmic rays.
\newblock \href{https://arxiv.org/abs/astro-ph/0107230}{{\em arXiv preprint}, 2001.}

\bibitem{Strong1998}
A.~W. Strong and I.~V. Moskalenko.
\newblock Propagation of Cosmic-Ray Nucleons in the Galaxy.
\newblock \href{https://doi.org/10.1086/306470}{{\em The Astrophysical Journal}, 509:212–228, 1998.}

\bibitem{Giacinti_2015}
G.~Giacinti, M.~Kachelrieß, and D.~V. Semikoz.
\newblock Escape model for Galactic cosmic rays and an early extragalactic transition.
\newblock \href{https://doi.org/10.1103/PhysRevD.91.083009}{{\em Physical Review D}, 91(8):083009, 2015.}


\bibitem{Effenberger_2012}
F.~Effenberger, H.~Fichtner, K.~Scherer, and I.~Büsching.
\newblock Anisotropic diffusion of Galactic cosmic ray protons and their steady-state azimuthal distribution.
\newblock \href{https://doi.org/10.1051/0004-6361/201220203}{{\em Astronomy \& Astrophysics}, 547:A120, 2012.}

\bibitem{Evoli_2017}
C.~Evoli, D.~Gaggero, A.~Vittino, et~al.
\newblock Cosmic-ray propagation with DRAGON2: I. numerical solver and astrophysical ingredients.
\newblock \href{https://doi.org/10.1088/1475-7516/2017/02/015}{{\em Journal of Cosmology and Astroparticle Physics}, 2017(02):015, 2017.}

\bibitem{Cao_2023}
Z.~Cao, F.~Aharonian, Q.~An, et~al.
\newblock Measurement of Ultra-High-Energy Diffuse Gamma-Ray Emission of the Galactic Plane from 10TeV to 1PeV with LHAASO-KM2A.
\newblock \href{https://doi.org/10.1103/PhysRevLett.131.151001}{{\em Physical Review Letters}, 131(15):151001, 2023.}

\bibitem{Zhang_2023}
R.~Zhang, X.~Huang, Z.-H. Xu, S.~Zhao, and Q.~Yuan.
\newblock Galactic Diffuse $\gamma$-Ray Emission from GeV to PeV Energies in Light of Up-to-date Cosmic-Ray Measurements.
\newblock \href{https://doi.org/10.3847/1538-4357/acf842}{{\em The Astrophysical Journal}, 957(1):43, 2023.}

\bibitem{Amenomori_2021}
M.~Amenomori, Y.~W. Bao, X.~J. Bi, et~al.
\newblock First Detection of sub-PeV Diffuse Gamma Rays from the Galactic Disk: Evidence for Ubiquitous Galactic Cosmic Rays beyond PeV Energies.
\newblock \href{https://doi.org/10.1103/PhysRevLett.126.141101}{{\em Physical Review Letters}, 126(14):141101, 2021.}

\bibitem{abbasi2024observationcosmicrayanisotropysouthern}
R.~Abbasi, M.~Ackermann, J.~Adams, et~al.
\newblock Observation of Cosmic-Ray Anisotropy in the Southern Hemisphere with Twelve Years of Data Collected by the IceCube Neutrino Observatory.
\newblock \href{https://arxiv.org/abs/2412.05046}{{\em arXiv preprint}, 2024.}

\bibitem{Ackermann_2012}
M.~Ackermann, M.~Ajello, A.~Albert, et~al.
\newblock Anisotropies in the diffuse gamma-ray background measured by the Fermi LAT.
\newblock \href{https://doi.org/10.1103/PhysRevD.85.083007}{{\em Physical Review D}, 85(8):083007, 2012.}

\bibitem{prevotat2024energydependencekneecosmic}
C.~Prévotat, M.~Kachelrieß, S.~Koldobskiy, A.~Neronov, and D.~Semikoz.
\newblock Energy dependence of the knee in the cosmic-ray spectrum across the Milky Way.
\newblock \href{https://doi.org/10.1103/PhysRevD.110.103035}{{\em Physical Review D}, 110(10):103035, 2024.}

\bibitem{Jansson_2012}
R.~Jansson and G.~R. Farrar.
\newblock A New Model of the Galactic Magnetic Field.
\newblock \href{https://doi.org/10.1088/0004-637X/757/1/14}{{\em The Astrophysical Journal}, 757(1):14, 2012.}

\bibitem{Unger_2024}
M.~Unger and G.~R. Farrar.
\newblock The Coherent Magnetic Field of the Milky Way.
\newblock \href{https://doi.org/10.3847/1538-4357/ad4a54}{{\em The Astrophysical Journal}, 970(1):95, 2024.}

\bibitem{Mertsch_2020}
P.~Mertsch.
\newblock Test particle simulations of cosmic rays.
\newblock \href{https://doi.org/10.1007/s10509-020-03832-3}{{\em Astrophysics and Space Science}, 365(8):1–17, 2020.}

\bibitem{kuhlen2023}
M.~Kuhlen, V.~H.~M. Phan, and P.~Mertsch.
\newblock Diffusion of relativistic charged particles and field lines in isotropic turbulence.
\newblock \href{https://arxiv.org/abs/2211.05881}{{\em arXiv preprint}, 2023.}

\bibitem{Kudryashov2023}
V. ~O. Yurovsky  and I.~A. Kudryashov.
\newblock Anisotropic Cosmic Ray Diffusion Tensor in a Numerical Experiment.
\newblock \href{https://doi.org/10.3103/S1062873823702337}{{\em Bulletin of the Russian Academy of Sciences: Physics}, 87(7):1123–1126, 2023.}

\bibitem{yusifov2004galacticdistributionluminosityfunction}
I.~Yusifov and I.~Kucuk.
\newblock Galactic Distribution and the Luminosity Function of Pulsars.
\newblock \href{https://arxiv.org/abs/astro-ph/0405495}{{\em arXiv preprint}, 2004.}

\bibitem{borisov2025modulationgalacticcosmicray}
V.~D. Borisov, V.~O. Yurovsky, and I.~A. Kudryashov.
\newblock Modulation of the galactic cosmic ray spectrum in an anisotropic diffusion approach.
\newblock \href{https://arxiv.org/abs/2502.19062}{{\em arXiv preprint}, 2025.}

\bibitem{Panov2007}
A.~D. Panov, J.~H. Adams, H.~S. Ahn, et~al.
\newblock Energy spectra of abundant nuclei of primary cosmic rays from the data of ATIC-2 experiment: Final results.
\newblock \href{http://dx.doi.org/10.3103/S1062873809050098}{{\em Bulletin of the Russian Academy of Sciences: Physics}, 71(4):494–497, 2007.}

\bibitem{Ahn2010}
H.~S. Ahn, P.~Allison, M.~G. Bagliesi, et~al.
\newblock Discrepant hardening observed in cosmic-ray elemental spectra.
\newblock \href{https://doi.org/10.1088/2041-8205/714/1/L89}{{\em The Astrophysical Journal Letters}, 714(1):L89–L93, 2010.}

\bibitem{Garyaka2008}
A.~P. Garyaka, R.~M. Martirosov, S.~V. Ter-Antonyan, et~al.
\newblock All-particle primary energy spectrum in the 3–200 PeV region according to the GAMMA experiment data.
\newblock \href{https://doi.org/10.1088/0954-3899/35/11/115201}{{\em Astroparticle Physics}, 28(2):169–180, 2008.}
 	



\bibitem{Aartsen_2019}
M.~G. Aartsen, M.~Ackermann, J.~Adams, et~al.
\newblock Cosmic ray spectrum and composition from PeV to EeV using 3 years of data from IceTop and IceCube.
\newblock \href{https://doi.org/10.1103/PhysRevD.100.082002}{{\em Physical Review D}, 100(8):082002, 2019.}

\bibitem{Apel2012}
W.~D. Apel, J.~C. Arteaga-Velázquez, K.~Bekk, et~al.
\newblock The KASCADE-Grande experiment.
\newblock \href{https://doi.org/10.1016/j.nima.2010.03.147}{{\em Nuclear Instruments and Methods in Physics Research Section A}, 620(2-3):202–216, 2012.}

\bibitem{Grebenyuk2019}
V.~Grebenyuk, D.~Karmanov, I.~Kovalev, et~al.
\newblock Secondary cosmic rays
in the NUCLEON space experiment.
\newblock \href{https://doi.org/10.1016/j.asr.2019.06.030}{\em Adv. Space Res. 64, 2559 (2019).}

\bibitem{Apanasenko2001}
A.~V. Apanasenko, Y.~A. Fedorov, V.~A. Galkin, et~al.
\newblock Energy Spectra of Abundant Cosmic-Ray Components in the Energy Range 10 TeV to 1000 TeV Observed by the RUNJOB Experiment.
\newblock \href{https://doi.org/10.1086/432715}{{\em The Astrophysical Journal}, 547(1):L33–L36, 2001.}

\bibitem{novotný2025energyspectrummasscomposition}
V.~Novotný.
\newblock Energy spectrum and mass composition of cosmic rays from Phase I data measured using the Pierre Auger Observatory.
\newblock \href{https://arxiv.org/abs/2501.01736}{{\em arXiv preprint}, 2025.}

\bibitem{disciascio2022measurementenergyspectrumelemental}
G.~Di Sciascio.
\newblock Measurement of Energy Spectrum and Elemental Composition of PeV Cosmic Rays: Open Problems and Prospects.
\newblock \href{https://arxiv.org/abs/2202.11618}{{\em arXiv preprint}, 2022.}

\bibitem{Budnev_2020}
N.~M. Budnev, A.~Chiavassa, O.~A. Gress, et~al.
\newblock The primary cosmic-ray energy spectrum measured with the Tunka-133 array.
\newblock \href{https://doi.org/10.1016/j.astropartphys.2019.102406}{{\em Astroparticle Physics}, 117:102406, 2020.}

\bibitem{Koldobskiy_2021}
S.~Koldobskiy, M. ~Kachelrieß, A. ~Lskavyan, et~al.
\newblock Energy spectra of secondaries in proton-proton interactions.
\newblock \href{https://doi.org/10.1103/PhysRevD.104.123027}{\em Phys. Rev. D, vol. 104, no. 12, p. 123027, 2021.}

\bibitem{Ostapchenko2006}
S.~Ostapchenko.
\newblock QGSJET-II: Towards reliable description of very high energy hadronic interactions.
\newblock \href{https://arxiv.org/abs/hep-ph/0412332}{{\em Nuclear Physics B - Proceedings Supplements}, 151:143–146, 2006.}

\bibitem{Lipari_2018}
P.~Lipari and S.~Vernetto.
\newblock Diffuse Galactic gamma-ray flux at very high energy.
\newblock \href{https://doi.org/10.1103/PhysRevD.98.043003}{{\em Physical Review D}, 98(4):043003, 2018.}

\bibitem{Vernetto_2016}
S.~Vernetto and P.~Lipari.
\newblock Absorption of very high energy gamma rays in the Milky Way.
\newblock \href{https://doi.org/10.1103/PhysRevD.94.063009}{{\em Physical Review D}, 94(6):063009, 2016.}



\end{thebibliography}
\end{document}